\documentclass{aa}

\usepackage[]{graphics}
\usepackage[]{psfig}
\usepackage[]{epsfig}
\usepackage[]{longtable}
\usepackage[]{times}
\include{epsf}
\begin{document}

\title{The distance to Hydra and Centaurus 
  from
surface brightness fluctuations: Consequences for the Great Attractor model\thanks{Based on observations obtained at the European 
Southern Observatory,
    Chile (Observing Programme 65.N--0459(A).}}

\author {Steffen Mieske \inst{1,2} \and Michael Hilker \inst{1} \and Leopoldo
Infante \inst{2}}

\offprints {S.~Mieske}
\mail{smieske@astro.uni-bonn.de}

\institute{
Sternwarte der Universit\"at Bonn, Auf dem H\"ugel 71, 53121 Bonn, Germany
\and
Departamento de Astronom\'\i a y Astrof\'\i sica, P.~Universidad Cat\'olica,
Casilla 104, Santiago 22, Chile
}

\date {Received 2 July 2004 / Accepted 4 March 2005}

\titlerunning{Distance to Hydra and Centaurus}

\authorrunning{S.~Mieske et al.}

\abstract{We present $I$-band Surface Brightness Fluctuations (SBF) measurements for 
16 early 
type galaxies (3 giants, 13 dwarfs) in the central
region of the Hydra cluster, based on deep photometric data in 7 fields obtained
with VLT FORS1.
From the SBF-distances to the galaxies in our sample we estimate the distance of the 
Hydra cluster to be 41.2 $\pm$ 1.4 Mpc ($(m-M)=$33.07 $\pm$ 0.07 mag). Based on an improved correction
for fluctuations from undetected point sources, we revise
the SBF-distance to the Centaurus cluster 
from Mieske \& Hilker (\cite{Mieske03b}) upwards by 10\% to 45.3 $\pm$ 2.0 Mpc ($(m-M)=$33.28 $\pm$ 0.09 mag).
The relative distance modulus of the two clusters then is $(m-M)_{\rm Cen}-(m-M)_{\rm Hyd}=0.21
\pm 0.11$ mag. 
With $H_{\rm 0}=$ 72 $\pm$ 4 km~s$^{-1}$ Mpc$^{-1}$, we estimate a positive
peculiar velocity of 1225 $\pm$ 235 km~s$^{-1}$ for Hydra and 210 $\pm$ 295 km~s$^{-1}$ for the Cen30 component of Centaurus. Allowing for a thermal velocity dispersion of 200 km~s$^{-1}$, this rules out a common peculiar flow velocity for both clusters at 98\% confidence. We find that the $9\times 10^{15} M_{\sun}$``Great Attractor'' from the flow study of Tonry et al.~(\cite{Tonry00}) at a distance of $\simeq$ 45 Mpc can explain the observed peculiar velocities if shifted about 15$\degr$ towards the Hydra cluster position. 
Our results are inconsistent at 94\% confidence with a scenario where the Centaurus cluster is identical to the GA.
In order to better restrict partially degenerate Great Attractor parameters like its mass and distance, a recalculation of the local flow model with
updated distance information over a larger area than covered by us would be needed.}

\maketitle


\section{Introduction}
\label{Introduction}
\subsection{Peculiar velocities of galaxies}
In the past few years, a lot of effort has been put into a precise
determination of cosmological parameters. Investigations like the HST Key
Project (Freedman et al.~\cite{Freedm01}),
the Sloan Digital Sky Survey (SDSS) (Abazajian et al.~\cite{Abazaj03}) or
the WMAP mission 
(Bennett et al.~\cite{Bennet03}, Spergel et al.~\cite{Sperge03}) mark the beginning
of a precision era in observational cosmology. The accuracy in determining the Hubble constant $H_{\rm 0}$
is approaching 5\%, most values derived recently by large surveys
are consistent with $H_{\rm 0}=72 \pm 4$ km~s$^{-1}$ Mpc$^{-1}$. This enormous improvement in precision 
has the consequence that 
  deviations from an undisturbed Hubble flow can also be
determined to a higher precision. This is of special importance for studying
the matter distribution in the nearby universe, as peculiar
velocities
caused by inhomogeneous matter distribution can be significant compared to
the Hubble flow.\\
There are three main methods that have been applied to estimate the distances
and peculiar
velocity field in the nearby universe:\\
1. The surface brightness fluctuation (SBF) method (e.g. Tonry \&
Schneider~\cite{Tonry88}, Jacoby et al.~\cite{Jacoby92}, Tonry et
al. ~\cite{Tonry01}). Tonry et al.~(\cite{Tonry00}), in the following T00, have
performed a study of the local flows using SBF distances obtained in the
$I$-band of about 300 early-type galaxies with $cz<4000$ km~s$^{-1}$. Apart from finding a
well defined flow towards the Virgo cluster, attributed to a ``Virgo
Attractor'' at about 17 Mpc distance, their model fitting favours an additional ``Great Attractor'' (GA)
of 9$\times$ 10$^{15}$ M$_{\sun}$
at a distance of 43 $\pm$ 3 Mpc in the direction of the Hydra-Centaurus
region. According to the SBF flow-model by
T00, the gravitational pull exerted by the GA
leads to a Local Group peculiar velocity of 300 $\pm$ 200 km~s$^{-1}$ with respect
to the Cosmic Microwave Background (CMB).\\
 2. The Fundamental Plane (FP) method (e.g. Dressler et al.~\cite{Dressl87}, 
Blakeslee et al.~\cite{Blakes02}, Bernardi et al.~\cite{Bernar03}).
Important early evidence for the presence of a Great Attractor towards the
Hydra-Centaurus region had come from an FP analysis performed by
Lynden-Bell et al. (\cite{Lynden88}). Later, several studies indicated that
the peculiar velocity of the Local Group with respect to the CMB could be only
partially explained by the attraction of the GA suggesting an additional
flow component towards a more distant attractor
(e.g. Willick~\cite{Willic90}, Hudson~\cite{Hudson94}, Staveley-Smith et
al.~\cite{Stavel00},
Lauer \& Postman~\cite{Lauer94}). Not all of these different studies found flow components towards the same direction. For instance, the large-scale flow detected by Lauer \& Postman was in a direction perpendicular to most other studies.
The ``Streaming motion of Abell clusters'' (SMAC)
survey (e.g. Smith et al.~\cite{Smith00}, \cite{Smith01}, \cite{Smith04} and 
Hudson et al.~\cite{Hudson04}) is the most recent example for an application
of the FP method to obtain peculiar velocities. In the SMAC survey, 
FP distances to 56 Abell galaxy 
clusters with $cz<12000$ km~s$^{-1}$ are presented. In Hudson et
al. (\cite{Hudson04}), a flow analysis based on these peculiar velocities is
performed, resulting in the detection of a bulk flow of amplitude
687 $\pm$ 203 km~s$^{-1}$ towards l=260$\degr$, b=0$\degr$ (SGL=170$\degr$, SGB=-57$\degr$), out
to 120 h$^{-1}$ Mpc.\\ 
3. The Tully-Fisher method (Tully \& Fisher~\cite{Tully77}). Most of the flow studies using this method (e.g. Aaronson et al.~\cite{Aarons82} and~\cite{Aarons86}, Mathewson et al.~\cite{Mathew92}, Shaya, Tully \& Pierce~\cite{Shaya92}, Courteau et al.~\cite{Courte93}, Dale et al.~\cite{Dale99}, Willick~\cite{Willic99}) detect a bulk flow of similar amplitude and direction to the one derived in the SMAC (Hudson et al.~\cite{Hudson04}) and find no indications for a convergence of this flow within $\simeq$ 100 $h^{-1}$ Mpc. There are also some studies which find no or much smaller bulk flows, e.g. Giovanelli et al.~(\cite{Giovan99}) or Courteau et al.~(\cite{Courte00}).\\
Distances derived for early type galaxies 
with the SBF-method and FP analysis are compared in Blakeslee et
al. (\cite{Blakes02}), showing an overall
good agreement between both methods. 
The distance differences between both methods are mostly statistical and not correlated with galaxy
properties, except for a mild dependence of the FP distance on the Mg$_2$ index.\\
\subsection{Peculiar velocities towards Hydra-Centaurus}
In Mieske \& Hilker (\cite{Mieske03b}), in the following MH, $I$-band SBF measurements for 15 early
type
Centaurus cluster galaxies were presented. A mean distance of 41.3 $\pm$ 2.1
Mpc (33.08 $\pm$ 0.11 mag)
towards the entire sample was derived. This distance implied only a few hundred km~s$^{-1}$ peculiar velocity against the Hubble flow for the Cen30 component of Centaurus. Anticipating Sects.~\ref{cencor} and~\ref{H0}, we note that in this paper the Centaurus cluster distance is revised upwards by about 10\%, implying an even smaller peculiar velocity. Tonry et
al. (\cite{Tonry00} and \cite{Tonry01}) had derived a smaller distance to
Centaurus by more than 0.5 mag, which in turn resulted in peculiar
velocities of about 1000 km~s$^{-1}$. It was stated in MH
that the lower peculiar velocities found for the Centaurus cluster with our higher quality data either imply
a smaller mass for the Great Attractor and/or an almost tangential infall
of Centaurus into the GA. Another possible interpretation of our results is that the Centaurus cluster actually {\it is} the Great Attractor and therefore at rest with respect to the CMB.\\
Adding a new distance$-$redshift pair
for the nearby Hydra cluster should allow to better discriminate between these possibilities, as the projected position of the GA determined by T00 is  
between Centaurus and Hydra.\\
Note that Hudson et al. (\cite{Hudson04}) show that the bulk flow -- in
which the Local Group participates -- points towards the approximate
direction of the Hydra-Centaurus region but extends well beyond the proposed position of the Great Attractor. 
They argue that this flow might mainly be caused by the
attractive force of the Shapley Super Cluster ($8000<cz<18000$
km~s$^{-1}$, see for example Quintana et al.~\cite{Quinta95}). However, due to the
 relatively sparse sampling close to the proposed Great
Attractor region, Hudson et al. cannot rule out that an additional massive 
attractor exists close to the Hydra-Centaurus region.\\
In our
present investigation, we will only check the GA model by T00 and
not test for any additional bulk flow. This is because the projected
positions of the Hydra and Centaurus
cluster are separated by only 30$\degr$ and the proposed position of the GA
is between the two clusters.
Due to this proximity, any other large scale streaming flow should have a negligible effect on their
relative velocities.\\
Note that here and in the following, the term ``Great Attractor'' (GA) refers to the mass overdensity whose location was estimated by T00 using the SBF-method. Its projected position between the Centaurus and Hydra clusters is at more than 40 degrees lower galactic longitude than the ``Great Attractor region'' from the studies of the galaxy density and peculiar velocity field in the Zone of Avoidance (ZOA) (e.g. Woudt et al.~\cite{Woudt03}, Kolatt et al.~\cite{Kolatt95}), specifically the massive Norma cluster. This might partially be because T00 did not observe galaxies in the Zone of Avoidance. However, it also reflects the clumpiness of matter distribution in the nearby universe. There might not be just one
major ``Great Attractor'', but the SBF-GA and ZOA-GA may be distinct substructures of a generally overdense filamentary region (e.g. Fairall \& Lahav~\cite{Fairal04}).
\subsection{Contents of this paper}
 In this paper we present new SBF distances to 16 early type Hydra 
galaxies -- 3 giants and 13 dwarfs. Compared to the SBF-measurement of Centaurus galaxies from MH, we will use an improved method of estimating the contribution of fluctuations from undetected point sources. Consequently, we apply these improved methods also to our previous Centaurus SBF-measurements and will present in this paper a revised distance set to Centaurus.\\
The aim of this paper is to improve the
distance precision 
to the Hydra cluster and calculate the relative 
distance between the Hydra and Centaurus cluster, taking advantage of the fact that the Hydra and Centaurus imaging data were obtained with the same instrument (VLT/FORS1). With the Hydra {\it and}
Centaurus distance in hand, some of Tonry's flow model parameters (Tonry et
al.~\cite{Tonry00}) for the Great Attractor can be checked.\\
This paper is structured as follows: In section \ref{datareduction}, 
the data and their 
reduction are described, including the revision of Centaurus distances.
Section \ref{results} shows the results of the SBF distance estimates for Hydra and Centaurus. The 
results are discussed with respect to the Great Attractor model
and compared with literature distances in section \ref{discussion}. 
We finish this paper 
with the conclusions in section \ref{conclusions}.
\section{The data}
\label{datareduction}
\begin{figure}
\begin{center}
\psfig{figure=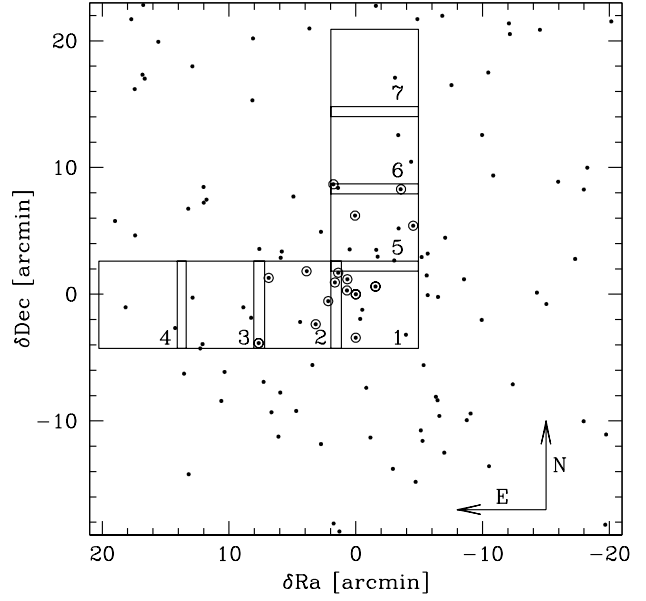,width=8.6cm}
\end{center}
\caption[]{\label{mapofhyd}Map of the central Hydra cluster, with distances relative 
to NGC 3311. Large squares are the observed VLT fields, the 
field number is indicated in the lower right corner of each square. 
Dots are all Hydra cluster member galaxies from the spectroscopic study
of Christlein \& Zabludoff (\cite{Christ03}).
Galaxies marked with large circles are the ones for 
which we present new SBF measurements in this paper. The cluster members without
SBF measurement are 11 late-type and 8 early-type galaxies.}
\end{figure}

The imaging data for Hydra have been obtained in service mode at the Very Large 
Telescope (VLT) of the European Southern Observatory,
Chile (Observing Programme 65.N--0459(A)), using UT 1 with the instrument FORS1 in imaging 
mode. Seven $7\times7$' fields in the central Hydra cluster have been observed in 
Johnson $V$ and $I$ pass-bands. The seeing ranged between 0.6 and 0.7$\arcsec$. The total 
integration time was 1500 seconds for the $V$ exposures, divided up into 4 dithered 
single exposures, and 3000 seconds for the $I$ exposures, divided up into 9 dithered 
single exposures. Fig.~\ref{mapofhyd} shows a map of the central Hydra cluster 
with the observed fields and indicating the cluster galaxies. Table~\ref{obs} gives the 
coordinates and photometric calibration coefficients of the observed fields. Table~\ref{galprob} gives the photometric 
properties and coordinates of the 16 investigated cluster galaxies. They span a magnitude
range of $10<V<18.5$ mag, corresponding to approximately  $-23<M_{\rm V}<-14.5$. The galaxies
investigated were selected from the radial velocity catalog of Christlein \& Zabludoff 
(\cite{Christ03}), requiring that they are early types and have radial velocities within
the Hydra cluster range of $2000<v_{\rm rad}<6000$ km~s$^{-1}$. The catalog has a complete spatial
coverage over the fields investigated by us. Its faint magnitude limit coincides
roughly with the faint magnitude limit for SBF measurements.\\
In the 7 fields, there are 8 additional early-type galaxies whose radial velocities
correspond to the Hydra cluster but which were not investigated.
One of them was too faint to detect a significant SBF signal. The other 7
galaxies showed pronounced boxy or disky residuals after subtracting an elliptical light
model with scale sizes of only a few PSF-FWHM.
\begin{table*}
\begin{center}
\begin{tabular}{lllllrlll}
Field & RA [2000] & Dec [2000] & ZP$_{\rm I}$ & ZP$_{\rm V}$ & CT$_{\rm I}$ & CT$_{\rm V}$ & k$_{\rm I}$ & k$_{\rm V}$\\\hline 
1 & 10:36:36.0 & -27:32:50 & 26.629  & 27.477  & -0.02  & 0.05  & 0.090  & 0.160  \\
2 & 10:37:03.4 & -27:32:50 & 26.643  & 27.529  & -0.02  & 0.05  & 0.090  & 0.160  \\
3 & 10:37:30.7 & -27:32:50 & 26.643  & 27.529  & -0.02  & 0.05 & 0.090  & 0.160  \\
4 & 10:37:58.6 & -27:32:50 & 26.543  & 27.529  & -0.02  & 0.05 & 0.090  & 0.160  \\
5 & 10:36:36.0 & -27:26:45 & 26.665  & 27.532  & -0.02  & 0.05 & 0.090  & 0.160  \\
6 & 10:36:36.0 & -27:20:39 & 26.665  & 27.532  & -0.02  & 0.05 & 0.090  & 0.160  \\
7 & 10:36:36.0 & -27:14:27 & 26.665  & 27.532  & -0.02  & 0.05 & 0.090  & 0.160  \\
\end{tabular}
\end{center}
\caption[]{\label{obs}Central coordinates and photometric calibration coefficients 
for the 7 VLT FORS1 fields as indicated in Fig.~\ref{mapofhyd}.}
\end{table*}
\begin{table*}
\begin{center}
\begin{tabular}{llllrrrl}
Nr.& Field & RA [2000] & Dec [2000] & V$_0$ [mag] & (V-I)$_0$ [mag] & $v_{rad}$ [km~s$^{-1}$] & Type \\\hline 
 258    & 1     &  10:36:50.1 & -27:30:46 & 17.57       & 1.02 & 5251      & dE,N \\
 359    & 1     &  10:36:49.0 & -27:30:00 & 17.44       & 1.04 & 4556     & dE \\
 334    & 1     &  10:36:45.8 & -27:31:24 & 17.65        & 1.03 & 4225     & dE,N  \\
 357    & 1     &  10:36:45.7 & -27:30:31 & 17.30       & 0.93 & 3815     & dE  \\
 140    & 1     &  10:36:42.7 & -27:35:08 & 16.15       & 1.05 & 3554     & dS0,N \\
 N3311& 1     &  10:36:42.7 & -27:31:42 & 10.90       & 1.15 & 3713     &
 S0(2) [E+2] \\
 N3309& 1     &  10:36:35.7 & -27:31:05 & 11.90       & 1.21 & 4068     & E1 [E3]  \\
 482   & 2     &  10:37:13.7 & -27:30:25 & 16.73        & 1.06 & 4439     & dE,N \\
 421   & 2     &  10:37:00.1 & -27:29:53 & 18.11       & 1.04 & 5626     & dE,N\\
 123    & 2     &  10:36:57.0 & -27:34:04 & 18.51       & 1.03 & 2993     & dE   \\
 172    & 2     &  10:36:52.5 & -27:32:15 & 16.28       & 1.07 & 3133     & dE \\
 252    & 3     &  10:37:17.3 & -27:35:34 & 17.19       & 1.08 & 3780     & dE,N  \\
 358     &  5  &   10:36:43.0 & -27:25:30 & 17.69 & 1.06 & 3936 & dE,N \\
 N3308 & 5     &  10:36:22.3 & -27:26:17 & 12.10       & 1.28 & 3537     & SB0(2) [SAB(s)0-] \\
 322  & 6 & 10:36:50.7 & -27:23:01 & 18.24 & 1.02 & 4306 & dE \\
150  & 6 & 10:36:26.8 & -27:23:25 & 15.99 & 1.07 & 4158 & dE,N \\
\end{tabular}
\end{center}
\caption[]{\label{galprob}Coordinates and photometric properties of the investigated 
galaxies of the Hydra cluster. 
The galaxies are ordered by field-number, and within the same field by 
right ascension. The field number refers to the fields indicated in Fig.~\ref{mapofhyd}. 
Photometry is taken from this paper. $(V-I)$ is from the region where SBF are measured, V$_0$ is the total magnitude of the galaxy derived from a curve-of-growth analysis.
Galaxy numbers, coordinates and radial velocities are from the catalog of 
Christlein \& Zabludoff (\cite{Christ03}),
except for the NGC galaxy numbers. Galaxy types are according
to the morphology on our Hydra images except for the three NGC galaxies, for which the type
is from Richter et al. (\cite{Richte89}).}
\end{table*}
\subsection{Data reduction before SBF measurement}
The reduction steps before SBF measurement are 
bias and flat-field correction, image combining, cosmic
ray removal, photometry of the investigated galaxies and standard star
calibration as outlined in MH.
To correct for galactic reddening and 
absorption, we used the values from Schlegel et al. (\cite{Schleg98}), who give $A_I=0.154$ and 
$E(V-I)=0.110$ for the coordinates of the Hydra cluster.\\
\subsection{SBF measurement}
\label{sbfmeas}
For obtaining $(m-M)$ with SBF, one must {\it measure} the 
apparent fluctuation magnitude $\overline{m}_{\rm I}$ and {\it derive} the absolute 
fluctuation magnitude $\overline{M}_{\rm I}$ from
$(V-I)_0$. For $(V-I)>1.0$, the calibration by Tonry et al. (\cite{Tonry01}) is used:\\
\begin{equation}
\overline{M}_{\rm I}=-1.74 + 4.5 \times ((V-I) - 1.15)\;{\rm mag}
\label{sbfrel}
\end{equation}
For $(V-I)<1.0$, we adopt a shallower slope of 2.25 instead of 4.5. This is because of the
discrepancy of model predictions in that colour range (see Mieske et al. \cite{Mieske03a}), for which some
authors (e.g. Worthey \cite{Worthe94}) 
predict a continuation of the steep slope while others (e.g. Liu et al. 
\cite{Liu00}) predict a flatter slope.
A cosmic scatter of 0.10 mag is assumed for the calibration (see MH), adding
in quadrature to the error contribution arising from the uncertainty of $(V-I)$. From
comparing the colour measured for the same galaxy in two adjacent fields (see Fig.~\ref{vimi}), 
we adopt $\Delta (V-I)=$ 0.015 mag as the colour uncertainty.\\
The steps performed to measure $\overline{m}_{\rm I}$ are (see MH):\\
Model the galaxy, subtract the model,  detect and subtract contaminating sources. Model and 
subtract again, divide by square root of model, cut out image portion for SBF measurement, mask
out contaminating sources. Calculate the power spectrum (PS), divide PS by
the fraction of not-masked pixels. Obtain the azimuthal average. Fit function of 
the form
\begin{equation}
\label{azimut}
P(k)=PSF(k)\times P_{\rm 0}+P_{\rm 1}
\end{equation}
to the azimuthally averaged PS. PSF(k) is the undistorted PS of the PSF normalised to
unity at k=0. $P_{\rm 0}$ is the amplitude 
of the pixel-to-pixel surface brightness fluctuations. $P_{\rm 1}$ is the white noise component.
To disregard the effect of large scale residuals from the model subtraction, low wavenumbers
are excluded from the PS fit. It holds for $\overline{m}_{\rm I}$:
\begin{equation}
\label{mbarI}
\overline{m}_{\rm I}=-2.5*log(P_{\rm 0}) + ZP - \Delta {\rm sim}- A_I - \Delta k + \Delta GC
+ \Delta {\rm BG}
\end{equation}
with $ZP$ being the photometric zero point including exposure time. $A_I$ is 
the foreground absorption, 
$\Delta k=z\times 7$ the k-correction for SBF in the $I$-band (Tonry et al. \cite{Tonry97}). $\Delta {\rm sim}$ is the correction of the
window function effect, see 
Sect.~\ref{bias}.\\
$\Delta {\rm BG}$ is the relative contribution of the background variance to the SBF signal. For its calculation we use a background field imaged in the same observing run, about 3 degrees away from the cluster center. Here, the same detection parameters as for the masking of sources in the SBF fields are applied and the detected objects masked. The fluctuations in this cleaned image are measured. For each galaxy, $\Delta {\rm BG}$ is then estimated dividing the amplitude of these fluctuations by the mean intensity of the galaxy light in the region where SBF are measured. $\Delta {\rm BG}$ is typically of the order of 0.1 mag.\\
$\Delta GC$ is the contribution to the fluctuations caused by Globular Clusters (GCs) 
below the detection limit. It is calculated using equation~(15) from Blakeslee \&
Tonry~(\cite{Blakes95}), assuming
a turn-over magnitude (TOM) of $M_I=-8.46 \pm$ 0.2 
mag for the globular cluster luminosity function (Kundu \& Whitmore \cite{Kundu01}) and a width $\sigma=$ 1.2 mag. The specific frequency $S_N$ of the dwarfs' GC systems was estimated by counting the point sources detected in the dwarf galaxy image after subtraction of the elliptical model, assuming that 50\% of the GCs are detected (the completeness magnitude as determined from artificial star experiments roughly corresponds to the apparent turnover magnitude of the GCLF expected at Hydra's distance, see Table~\ref{resultsgcs} and Fig.~\ref{gclf}). The number of detected GCs per galaxy is generally below ten. Therefore, an average $S_N$ was determined for the five dwarfs in field 1 and the eight ones in the other fields. For field 1, which contains the extended halo GC population of the cD galaxy NGC 3311, we find a high average specific frequency of $S_N=11.5 \pm 2.8$, for the other fields we find $S_N=$2.3 $\pm$ 0.8. The apparent TOM of the GCLF was assumed to be equal for all dwarfs, derived from their mean SBF-distance.As $\Delta GC$ depends on the fraction of GCs that can be detected and hence correlates somewhat with galaxy distance, the mean $\Delta GC$ and mean SBF-distance of the dwarfs were determined iteratively, using the mean colour of all dwarfs. We note however, that this correlation is not very strong: changing the assumed TOM of the GCLF by 0.1 mag, which as we show in Sect.~\ref{results} is the uncertainty of the mean Hydra SBF-distance, changes $\Delta GC$ by less than 0.01 mag. To obtain $\Delta GC$ for a single dwarf, the mean $\Delta GC$ was changed according to the specific colour of that galaxy, resulting in a scatter of about 0.03 mag in $\Delta GC$ between the different dwarfs. The derived $\Delta GC$ was about 0.45 mag for the dwarfs in field 1 and generally below 0.1 mag for the other fields.\\
For NGC 3309 and NGC 3311, $\overline{m}_{\rm I}$ and $(V-I)_0$ were 
measured independently in three concentric rings, covering the range 
$8''<r<26''$ for NGC 3309 and $8''<r<32''$ for NGC 3311. $S_N$ and the apparent TOM are directly determined in the rings where SBF are measured, as summarised in Table~\ref{resultsgcs} and shown in Fig.~\ref{gclf}. For this fitting, we keep $\sigma$ fixed at 1.3 mag (Kundu \& Whitmore \cite{Kundu01}), with an error allowance of 0.15 mag. Due to the relatively low number of GCs present in the images and the somewhat brighter completeness magnitude as compared to the Centaurus data, the TOM of these GCLFs is only very poorly constrained. However, due to the bivariance of $\sigma$ and TOM, $\Delta GC$ is almost independent of both values. E.g. adopting $\sigma=1.1$ mag instead of $1.3$ mag results in a $0.5$ mag brighter TOM and 30\% lower $S_N$ for both NGC 3309 and 3311. This gives $\Delta GC=0.21$ mag for NGC 3311 and $\Delta GC=0.11$ mag for NGC 3309, lower by only 0.03-0.04 mag than the values obtained from using $\sigma=1.3$ mag.\\
The $S_N$ of NGC 3311 is lower by 1.7 $\sigma$ than the value of $S_N=15 \pm 6$ found by McLaughlin et al.~\cite{McLaug95}, but their study investigated radial distances between 0.5 and 3.5$\arcmin$, outside the area sampled by us. Our value is in qualitative agreement with previous findings that the local specific frequency decreases significantly in the innermost regions of cD galaxies (e.g. McLaughlin~\cite{McLaug94}, Forte et al.~\cite{Forte05}).\\
The most important error contribution of $(m-M)$ comes from $P_{\rm 0}$.
The error in $P_{\rm 0}$ is derived from the same Monte Carlo simulations as presented in
MH. It is adopted to be 0.26 mag for $V_0<16.6$ and 
0.42 mag for
$V>16.6$ mag. For NGC 3309 and NGC 3311, the error in $(m-M)$ is determined from the scatter
of the $(m-M)$ values in the three rings used for measuring SBF.
\subsubsection{The window function effect}
\label{bias}
\begin{figure*}[]
\begin{center}
\epsfig{figure=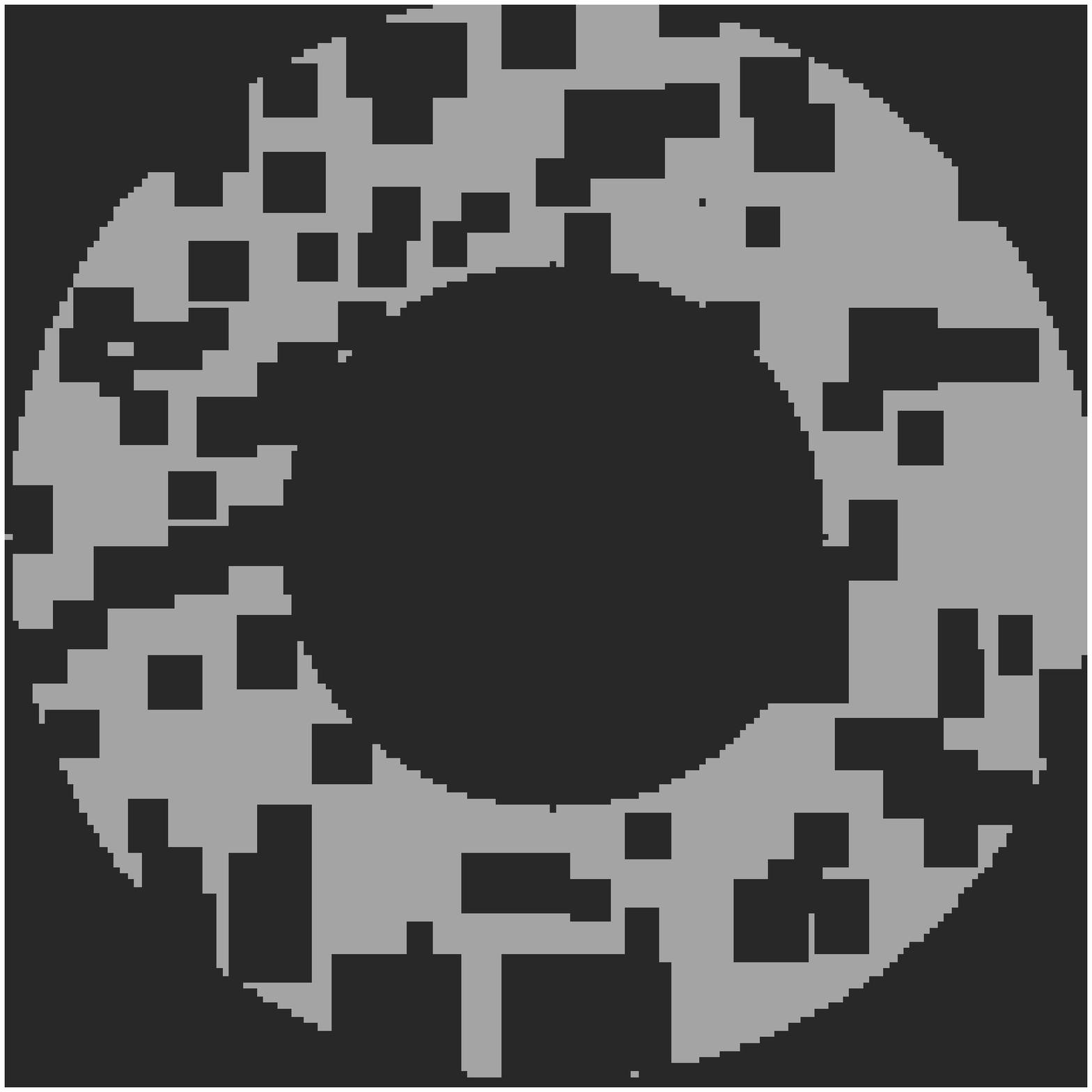,width=5.5cm}\hspace{0.1cm}
\epsfig{figure=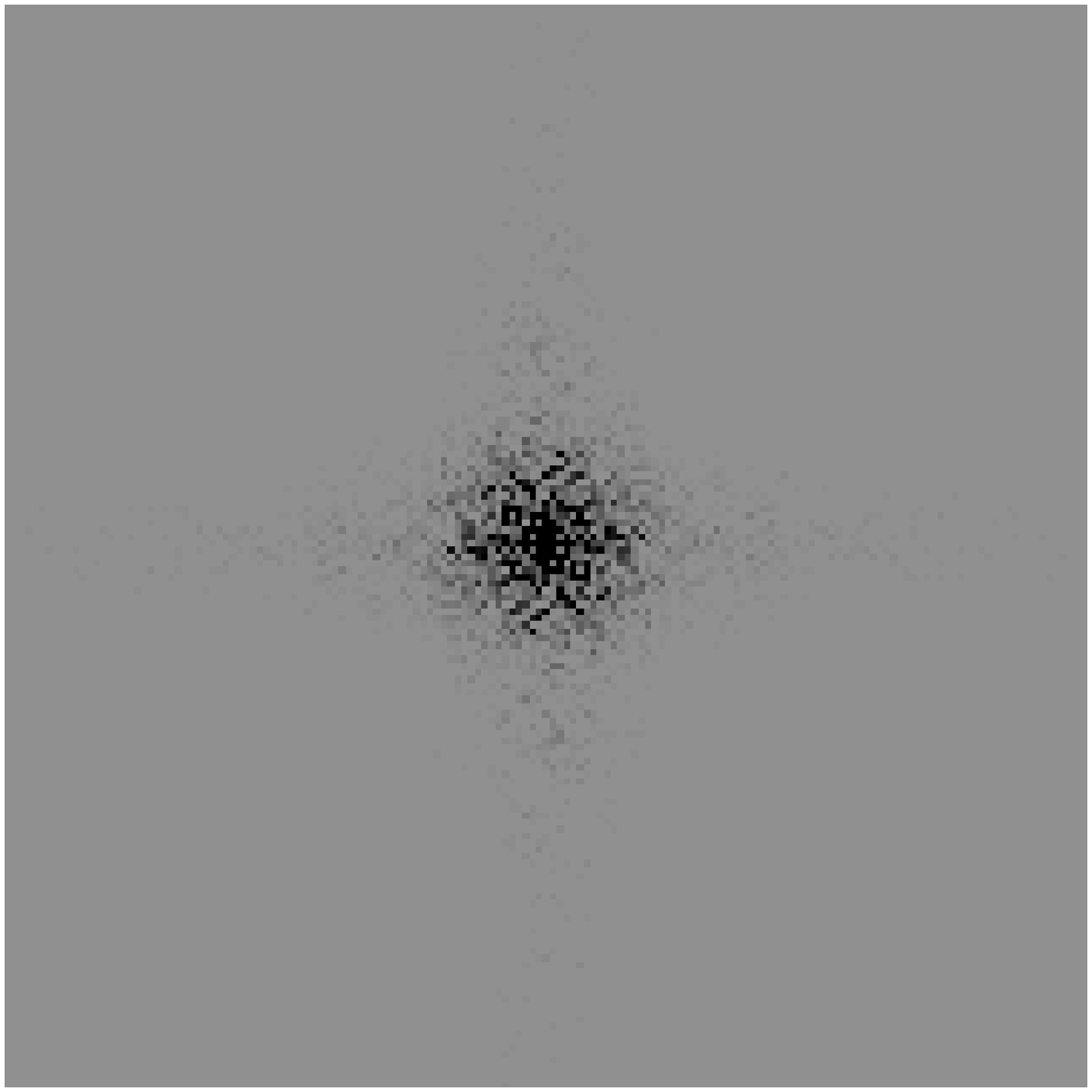,width=5.5cm}\hspace{0.1cm}
\epsfig{figure=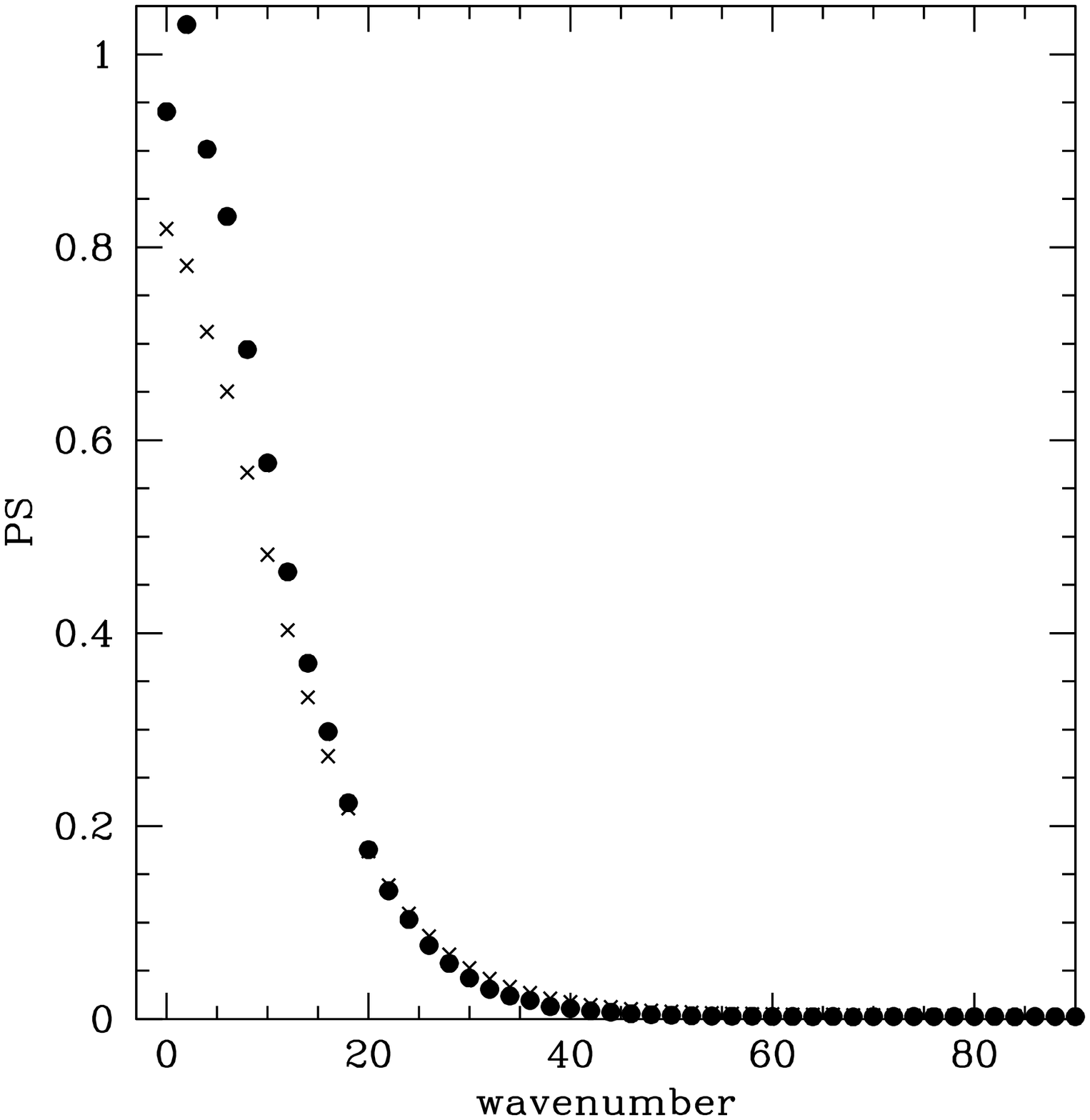,width=6.5cm}
\epsfig{figure=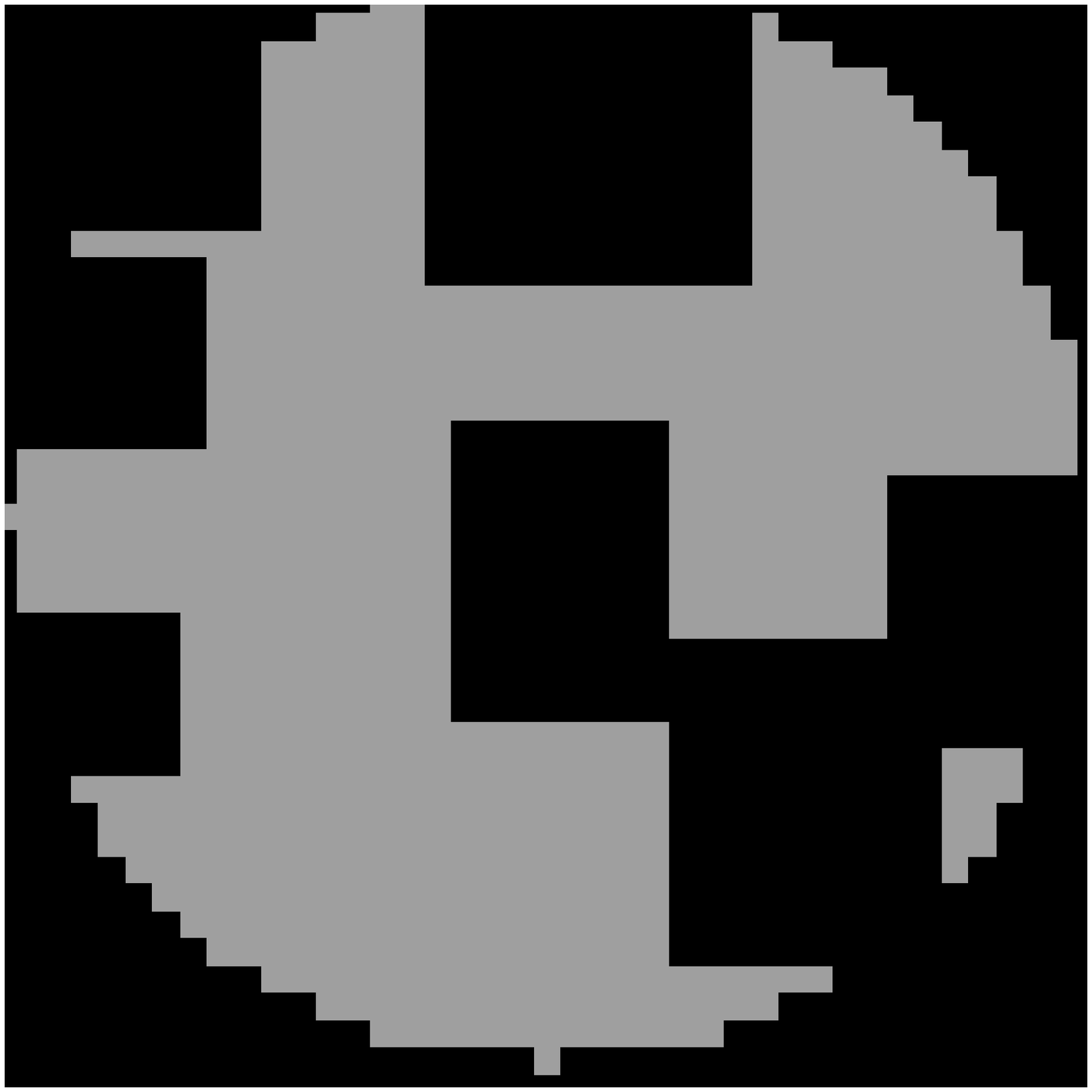,width=5.5cm}\hspace{0.1cm}
\epsfig{figure=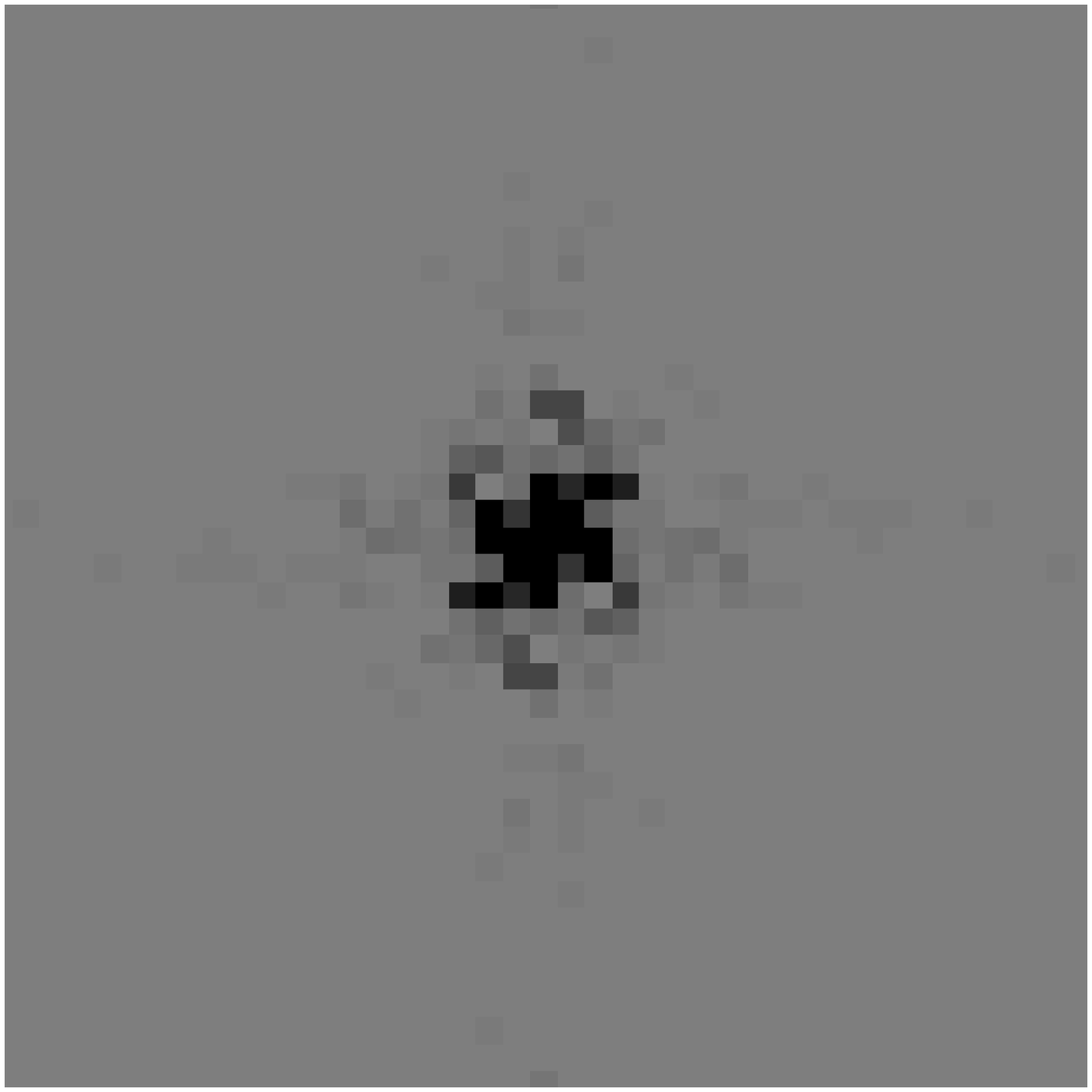,width=5.5cm}\vspace{0.1cm}
\epsfig{figure=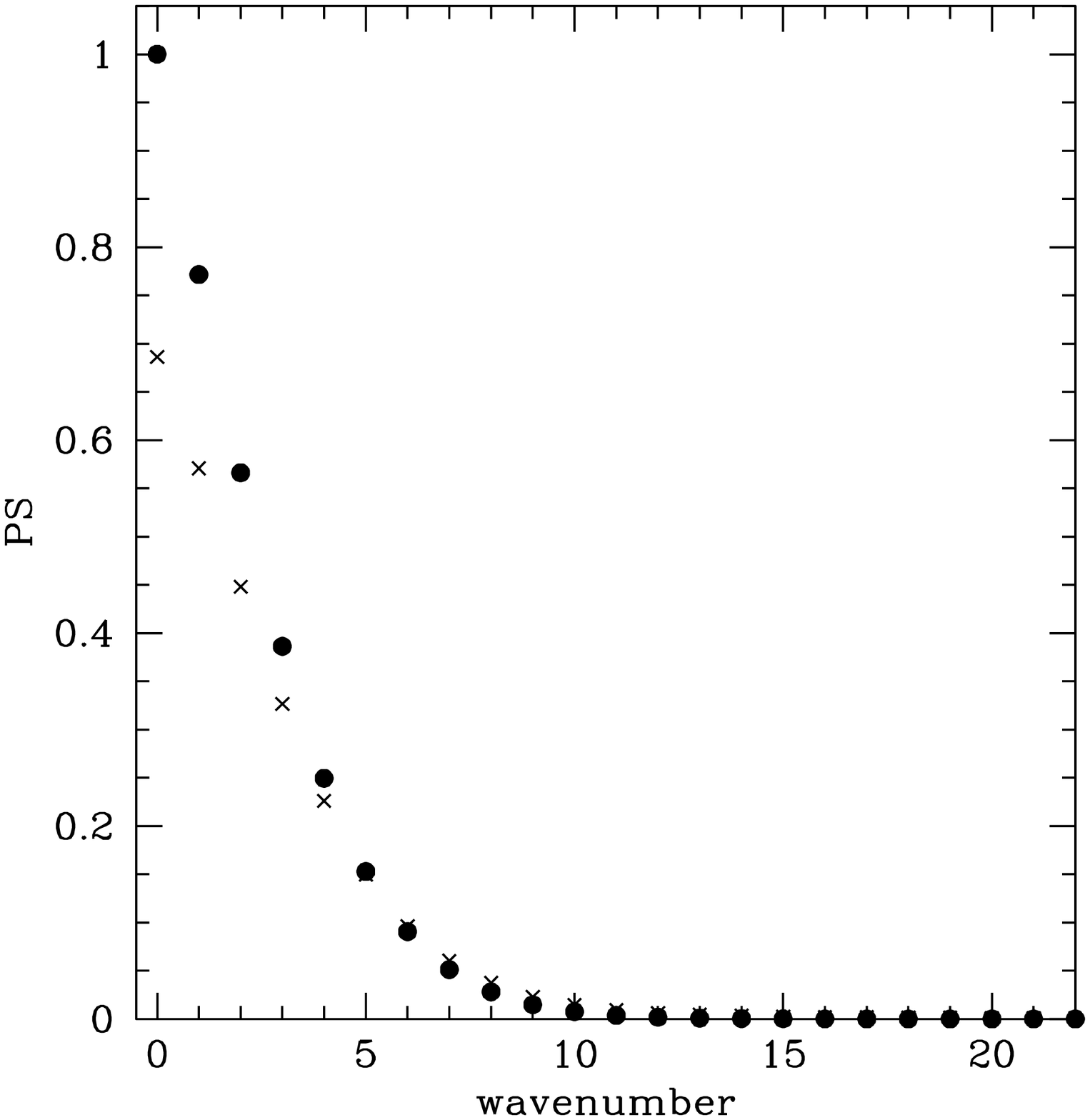,width=6.5cm}
\end{center}
\caption[]{\label{sbfbias}Plots illustrating the effect of the window
  function effect. {\it Upper panels:} The inner ring of NGC 3311. {\it
    Lower panels:} Galaxy 258 in field 1. {\it Left
panels:} mask image used to blend out contaminating sources and restrict
    the measurement region. Masked regions are black. {\it Middle panels:} Power spectrum of the mask
    on the left image. 
{\it Right panels:} Filled circles: power spectrum of
    an isolated star, normalised to unity at wavenumber k=0. Crosses: the same power spectrum convolved with the
    mask
power spectrum from the middle panels. The convolved power spectrum is
    lower than the non-convolved one by about 0.2 mag for NGC 3311 and about
    0.4 mag for galaxy 258,
    corresponding to the values derived with the SBF simulations
    (see Table~\ref{resultstab} and Sect.~\ref{bias}.)}
\end{figure*}

In the simulations of SBF measurements of dEs in nearby clusters (Mieske et al. \cite{Mieske03a}) 
it has been shown that at 0.5$\arcsec$ seeing the measured $\overline{m}_{\rm I}$ is about 0.15 $\pm$ 0.05 mag 
fainter than the simulated $\overline{m}_{\rm I}$. This difference is taken into account by including  $\Delta {\rm sim}$ in equation~\ref{mbarI}. 
The reason why this correction is needed is that in equation~\ref{azimut} we 
assume an undisturbed power spectrum
$PSF(k)$. Doing so implicitly neglects the fact that the FT of the seeing convolved SBF is in frequency space convolved
with the FT of the mask used to excise contaminating point sources. This results in a damping of the SBF amplitude at low wavenumbers (see Fig.~\ref{sbfbias}).\\
This can be corrected for by using the expectation power spectrum $E(k)=PSF(k) \otimes W(k)$ instead of $PSF(k)$ in equation~\ref{azimut}. Here, $W(k)$ is the power spectrum (PS) of the ``window function'' defined by the mask. This approach is used for example by the Tonry group (Tonry et al.~\cite{Tonry90}, \cite{Tonry97},
\cite{Tonry00}, ~\cite{Tonry01}, see also the review by Blakeslee et al.~\cite{Blakes99}) and 
in the publications by Jensen et al.~(\cite{Jensen98}, \cite{Jensen99}, \cite{Jensen01}). Although these authors
include $W(k)$, it is commonly stated that its effect on the expectation
power spectrum is small. For example, Jensen et al. (\cite{Jensen98}) remark that the
effect of 
the window function on the shape of the expectation
PS is ``minimal'' and that $E(k)$ is ``very nearly''
$PSF(k)$. There are also some authors that entirely neglect the window
function effect (e.g. Sodemann et al.~\cite{Sodema95}, Mei et
al.~\cite{Mei00}). 
Based on the results of
our simulations (Mieske et al.~\cite{Mieske03a})
and also the examples in
Fig.~\ref{sbfbias}, it is clear that the effect of $W(k)$ can amount to several tenths of
magnitude
and should always be taken into account.\\
Our approach to include $\Delta sim$ in equation~\ref{mbarI} occurs one step later in the measurement procedure than if using immediately $E(k)$. Nevertheless it is quite instructive as it illustrates well the 
overestimation in distance that can occur when not considering $W(k)$, see Fig.~\ref{sbfbias}. To calculate $\Delta sim$ for a given galaxy, first a portion of the same 
size as used for the galaxy SBF measurement is cut out of a much larger artificial SBF image carrying the fluctuation signal $P_{\rm 0,in}$. The latter image is created separately for each field by convolving simulated pixel-to-pixel SBF with a model of the PSF in the respective field. The excerpt is then multiplied by the mask image used in the real SBF measurement. The PS of this image is calculated, divided by the fraction
of not-masked pixels and $P_{\rm 0,out}$ is
determined according to equation~(\ref{azimut}). 
It holds $\Delta {\rm sim}=-2.5\times log(\frac{P_{\rm 0,out}}{P_{\rm 0,in}})$. The values derived are shown in Table~\ref{resultstab}, they are typically of the order 0.2 mag.
\subsection{Revision of the Centaurus cluster distance}
\begin{figure}[]
\begin{center}
\epsfig{figure=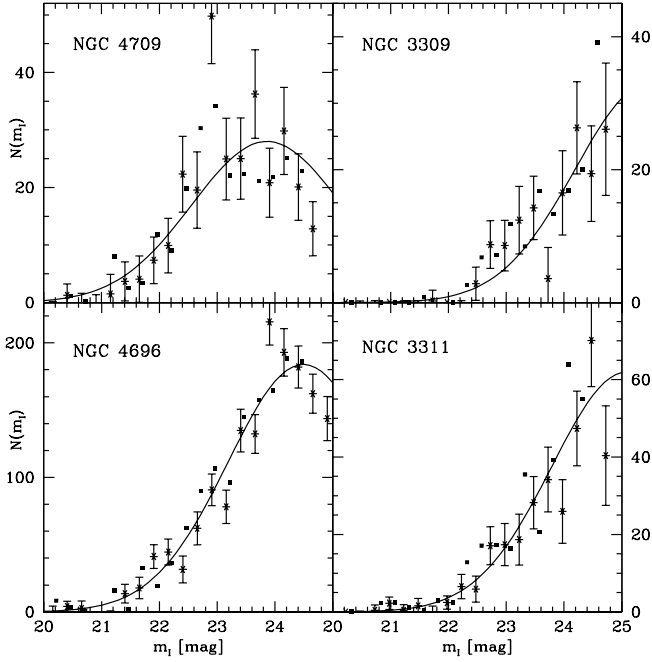,width=9cm}
\end{center}
\caption[]{\label{gclf}Incompleteness corrected globular cluster luminosity functions of the four indicated galaxies in the rings where SBF were measured. Asterisks with error bars are counts in the $I$-band, filled squares are counts in the $V$-band shifted to $I$ assuming $(V-I)=1.04$ from Kundu \& Whitmore (\cite{Kundu01}). The number counts are corrected for background contamination and shown up to the faintest magnitude bin for which the completeness was above 50\%. The solid lines are Gaussian fits, whose parameters are the mean of the values fit independently in $V$ and $I$. The derived turnover magnitudes and specific frequencies $S_N$ are given in Table.~\ref{resultsgcs}.}
\end{figure}

The derivation of $\overline{m}_{\rm I}$ from $P_0$ as given by equation~\ref{mbarI} in this paper includes the term $\Delta BG$ to account for residual variance present in a cleaned background field. $\Delta BG$ is measured directly in the background field for the Hydra galaxies, and found to be 0.10 mag on average, see Table~\ref{resultstab}. For the Centaurus cluster SBF-measurements in MH, these contributions were assumed to be negligible without a double check in the background field. Therefore, we have gone back to the Centaurus cluster background field and also measured directly the residual variance. It turns out to be slightly lower than for the Hydra cluster, but still between 0.01 and 0.20 mag. One reason for the underestimation in MH is that the average dwarf galaxy surface brightness assumed by us for the estimation of $\Delta BG$ was too bright. Furthermore, directly measured sky fluctuations also include the fluctuations caused by imperfect image quality, e.g. fringing, flat field effects. We have now included $\Delta BG$ in the derivation of $\overline{m}_{\rm I}$ for the Centaurus cluster galaxies, see the revised distance set in Table~\ref{revisedcen}.\\
A further change is that $\Delta_{\rm sim}$ is now calculated separately for each galaxy instead of assuming the same value 0.15 mag for all galaxies. Appropriate simulations analogous to the ones described in Sect.~\ref{bias} for the Hydra galaxies are used for that. The resulting values of $\Delta_{\rm sim}$ vary between 0.09 and 0.20 mag with a mean of 0.16 mag (see Table~\ref{revisedcen}).\\
We also double checked the specific frequency $S_N=$4 assumed for the GCLF of the Centaurus dwarfs. By counting the number of point sources detected on the model subtracted images (see Sect.~\ref{sbfmeas}), we estimate an average specific frequency of $S_N= 5.4$ $\pm$ 1.0, independent on location in the cluster. We therefore have recalculated $\Delta_{\rm GC}$ with this slightly higher $S_N$, see Table~\ref{revisedcen}. Furthermore, we assume now the same apparent TOM for each dwarf, based on the mean SBF distance of the entire sample. This removes the dependence of $\Delta_{\rm GC}$ on $(m-M)$, which occurred in MH because these two entities were calculated iteratively for each single galaxy rather than for the sample as a whole.\\
Finally, the globular cluster systems of NGC 4696 and NGC 4709 were re-analysed, see Table~\ref{resultsgcs} and Fig.~\ref{gclf}. Like for the Hydra galaxies, the GC detection in $V$ and $I$ was now made separately without any colour restriction, which resulted in a fainter completeness limit. Contribution of background sources to the GC counts was estimated from the background field. The revised specific frequencies are slightly higher than given in MH, the TOM is about 0.2 mag fainter for NGC 4696 and 0.3 mag fainter for NGC 4709. Also for these fits, $\sigma$ was kept fixed at 1.30 $\pm$ 0.15 mag. For NGC 4696, the galaxy with the highest number of GCs and faintest completeness limit in our sample, a fit with varying $\sigma$ gives $\sigma=1.24 \pm 0.12$ mag, consistent with the assumed fixed value.
\label{cencor}
\begin{table*}
\begin{center}
\begin{tabular}{r|llllllll}
Gal-Nr. & TOM [mag]& $\sigma$ & $I_{\rm cut}$ [mag] & $(m-M)_{\rm GC}$ & $(m-M)_{SBF}$&N$_{\rm GC, rings}$& $M_{\rm V, rings}$ & $S_{\rm N, rings}$\\\hline\hline
 NGC 3309 & 25.49 $\pm$ 0.70 &1.30 (fixed) & 24.70 & 33.95 $\pm$ 0.73 & 32.81 $\pm$ 0.13 & 430 $\pm$ 140 & $-$20.30 $\pm$ 0.2 & 3.5 $\pm$ 1.3\\
 NGC 3311 & 25.09 $\pm$ 0.59 &1.30 (fixed) & 24.70 &33.55 $\pm$ 0.63 & 33.09 $\pm$ 0.14 &  810 $\pm$ 300 & $-$20.60 $\pm$ 0.2 & 4.6 $\pm$ 1.9\\
 NGC 4696 & 24.49 $\pm$ 0.27 & 1.30 (fixed) & 25.05 &32.95 $\pm$ 0.34 & 33.14 $\pm$ 0.16 &  2390 $\pm$ 300 & $-$21.25 $\pm$ 0.2 &  7.3 $\pm$ 1.5\\
 NGC 4709 & 23.86 $\pm$ 0.26 & 1.30 (fixed) & 25.00 &32.32 $\pm$ 0.33 & 32.50 $\pm$ 0.15&  360 $\pm$ 60 & $-19.60$ $\pm$ 0.2 &    5.0 $\pm$ 1.3\\

\end{tabular}
\end{center}
\caption[]{\label{resultsgcs}Details of the globular cluster luminosity function (GCLF) fitting for the Hydra galaxies NGC 3309 and NGC 3311 and the Centaurus galaxies NGC 4696 and NGC 4709, performed in the rings where SBF were measured. A Gaussian with width $\sigma=1.3$ mag (Kundu \& Whitmore \cite{Kundu01}) with an error allowance for $\sigma$ of $\pm$ 0.15 mag is fit to the incompleteness corrected number counts in Fig.~\ref{gclf}. The error of the turnover magnitude (TOM) is the maximum of the fitting error and the difference in TOM when changing $\sigma$ to its lower and upper limit of 1.15 and 1.45 mag. $I_{\rm cut}$ is the limiting magnitude for the GCLF fitting, identical to the 50\% completeness limit for GC detection. An absolute turn-over magnitude of $M_I=-8.46 \pm$ 0.2 
mag is assumed (Kundu \& Whitmore \cite{Kundu01}). The GCLF-TOMs of the two Hydra galaxies are poorly constrained both because of a brighter cutoff magnitude and lower number of GCs compared to NGC 4696.}
\end{table*}
\section{Results}
The Hydra SBF-distances are shown in Table~\ref{resultstab}. The revised SBF-distances to the Centaurus cluster galaxies from MH are shown in Table~\ref{revisedcen}.
\begin{table*}
\begin{center}
\begin{tabular}{rrrrrrrrrrrr}
Nr. & Field & $P_0$ [ADU] & $P_1$ [ADU]& $ZP$ & $S/N$ & $\Delta {\rm sim}$ & $\Delta BG$ & $\Delta GC$ & $\overline{m}_{\rm I}$ & $(m-M)$ & $d$ [Mpc]\\\hline\hline
   258 & 1 & 4.06 $\pm$ 1.92 & 0.91 &  32.81 &  4.46 & 0.393 & 0.115 &0.43 &  31.16 $\pm$ 0.46 & 33.50 $\pm$ 0.47 &  50.19 $\pm$   11.0\\
   359 & 1 & 9.04 $\pm$ 4.24 & 1.44 &  32.81 &  6.28 & 0.305 & 0.122 &0.46 &  30.44 $\pm$ 0.46 & 32.68 $\pm$ 0.47 &  34.29 $\pm$  7.5\\
   334 & 1 & 7.51 $\pm$ 3.53 & 1.10 &  32.81 &  6.83 & 0.268 & 0.116 &0.45 &  30.67 $\pm$ 0.45 & 32.94 $\pm$ 0.47 &  38.76 $\pm$  8.5\\
   357 & 1 & 10.46 $\pm$ 4.92 & 1.06 &  32.81 &  9.87 & 0.210 & 0.065 &0.38 &  30.25 $\pm$ 0.45 & 32.83 $\pm$ 0.49 &  36.74 $\pm$  8.4\\ 
   140 & 1 & 3.66 $\pm$ 1.93 & 0.58 &  32.81 &  6.31 & 0.217 & 0.100 &0.47 &  31.52 $\pm$ 0.49 & 33.71 $\pm$ 0.51 &  55.15 $\pm$   13.0\\
N3309 & 1 & 3.18 $\pm$ 0.25 & 0.25 & 32.81 & 12.90 & 0.175 & 0.042 & 0.14 & 31.30 $\pm$ 0.09 &32.81 $\pm$ 0.13 & 36.47 $\pm$  2.1\\
N3311 & 1 & 3.29 $\pm$ 0.24  & 0.32 & 32.81 & 10.30 & 0.208 & 0.030 & 0.25 &31.33 $\pm$ 0.10 & 33.09 $\pm$ 0.14 &  41.42 $\pm$  2.6\\ 
 482  & 2 & 6.40 $\pm$ 3.00 & 1.93 &  32.82 &  3.32 & 0.204 & 0.241 &0.07 &  30.65 $\pm$ 0.42 & 32.96 $\pm$ 0.44 &  39.02 $\pm$    8.0\\
 421  & 2 & 4.61 $\pm$ 1.25 & 0.68 &  32.82 &  6.78 & 0.203 & 0.073 &0.08 &  30.83 $\pm$ 0.27 & 32.93 $\pm$ 0.30 &  38.54 $\pm$  5.3 \\
 123  & 2 & 4.08 $\pm$ 1.93 & 0.81 &  32.82 &  5.04 & 0.255 & 0.121 &0.08 &  31.02 $\pm$ 0.43 & 33.15 $\pm$ 0.44 &  42.69 $\pm$  8.8\\ 
 172  & 2 & 6.62 $\pm$ 3.11 & 1.34 &  32.82 &  4.94 & 0.108 & 0.135 &0.07 &  30.64 $\pm$ 0.42 & 32.85 $\pm$ 0.44 &  37.22 $\pm$  7.6\\
 252  & 3 & 4.18 $\pm$ 2.00 & 1.08 &  32.82 &   3.90 & 0.221 & 0.189 &0.08 &  31.07 $\pm$ 0.42 & 33.12 $\pm$ 0.42 &  42.05 $\pm$  8.3\\
 358  & 5 & 4.62 $\pm$ 2.17 & 0.94 &  32.85 &  4.91 & 0.119 & 0.145 &0.06 &  31.03 $\pm$ 0.42 & 33.16 $\pm$ 0.44 &  42.84 $\pm$  8.8\\
 N3308  & 5 &  2.24 $\pm$ 0.60 & 0.17 &  32.85 &  13.20 & 0.130 & 0.028 &0.24 &  31.88 $\pm$ 0.28 & 33.04 $\pm$ 0.30 &  40.63 $\pm$  5.7\\
 322  & 6 & 6.91 $\pm$ 3.25 & 0.78 &  32.84 &  8.86 & 0.248 & 0.142 &0.04 &  30.42 $\pm$ 0.42 & 32.73 $\pm$ 0.44 &  35.13 $\pm$  7.2 \\
 150  & 6 & 3.05 $\pm$ 0.82 & 0.46 &  32.84 &  6.63 & 0.238 & 0.109 &0.05 &   31.30 $\pm$ 0.27 & 33.40 $\pm$ 0.29 &  47.85 $\pm$  6.5\\\hline
& & & & & & & & &  &    {\bf 33.07 $\pm$ 0.07} & {\bf 41.20 $\pm$ 1.4}\\
\end{tabular}
\end{center}
\caption[]{\label{resultstab}Result of the SBF measurements for the investigated Hydra
cluster galaxies. The error in metric distance $d$ is the mean of the 
upper and lower distance error range
corresponding to the magnitude error in $(m-M)$. 
The columns $ZP$, $\Delta {\rm sim}$, $\Delta BG$, $\Delta GC$, $\overline{m}_{\rm I}$ and $(m-M)$ are given in magnitudes.
For the two giant galaxies NGC 3309 and NGC 3311, the results shown are the averaged means over the three rings
investigated. In the lowest row, the mean distance $d$ and the corresponding distance
modulus $(m-M)$ of all galaxies are given. }
\end{table*}
\begin{table*}
\begin{center}
\begin{tabular}{rrrrrrrrrrrr}
Nr. & Field & $P_0$ [ADU] & $P_1$ [ADU]& $ZP$ & $S/N$ & $\Delta {\rm sim}$ & $\Delta BG$ & $\Delta GC$ & $\overline{m}_{\rm I}$ & $(m-M)$ & $d$ [Mpc]\\\hline\hline
N4696 & 1 & 2.33 $\pm$ 0.04 & 0.488 &  32.78 &  4.78 & 0.090 & 0.090 &0.23& 31.840 $\pm$ 0.120 &33.14 $\pm$ 0.17 &  42.5 $\pm$   3.20 \\
  75 & 1,2 & 3.30 $\pm$ 0.35 & 0.86 &  32.77 &  3.84 & 0.178 & 0.079 &0.18 &  31.289 $\pm$ 0.430 &33.60 $\pm$ 0.45 &   52.5 $\pm$    11.0  \\
  61 & 1 & 2.94 $\pm$ 0.25 & 0.51 &  32.76 &  5.76 & 0.174 & 0.068 &0.26 &  31.456 $\pm$ 0.290 &33.23 $\pm$ 0.32 &   44.2 $\pm$   6.50\\ 
  70 & 1 & 3.56 $\pm$ 0.20 & 0.32 &  32.76 &  11.1 &  0.20 & 0.012 &0.34 &   31.260 $\pm$ 0.310 &32.58 $\pm$ 0.33 &  32.8 $\pm$   5.10\\
  52 & 1 & 2.81 $\pm$ 0.33 & 0.64 &  32.76 &  4.39 & 0.168 & 0.072 &0.22 &  31.473 $\pm$ 0.430 &33.47 $\pm$ 0.45 &  49.4 $\pm$    10.0\\  
  89 & 2 & 2.16 $\pm$ 0.13 & 0.29 &  32.78 &  7.45 & 0.150 & 0.020 &0.25 &  31.772 $\pm$ 0.290 & 33.50 $\pm$ 0.31 &  50.1 $\pm$   7.30  \\
N4709 & 3 & 1.90 $\pm$ 0.10 & 0.38 &  32.67 &   5.00 & 0.140 & 0.089 &0.06 &  31.623 $\pm$ 0.083 & 32.50 $\pm$ 0.15 & 31.6 $\pm$   2.20 \\
 124 & 3 & 6.28 $\pm$ 1.28 & 3.36 &  32.67 &  1.87 & 0.153 & 0.199 &0.13 &  30.561 $\pm$ 0.430 &  33.33 $\pm$ 0.57 &  46.3 $\pm$    12.0 \\
 123 & 3 & 5.38 $\pm$ 0.43 & 0.62 &  32.67 &  8.68 & 0.151 & 0.041 &0.19 &  30.594 $\pm$ 0.430 & 32.86 $\pm$ 0.45 &  37.3 $\pm$   7.70\\  
 121 & 3 & 2.65 $\pm$ 0.50 & 0.76 &  32.67 &  3.49 & 0.183 & 0.116 &0.24 &  31.454 $\pm$ 0.460 & 33.54 $\pm$ 0.47 &  51.1 $\pm$    11.0\\
 115 & 3 & 4.02 $\pm$ 0.59 & 1.34 &  32.67 &  3.00 & 0.145 & 0.169 &0.17 &  31.063 $\pm$ 0.430 & 33.49 $\pm$ 0.45 &  50.0 $\pm$    10.0\\  
 111 & 3 & 6.02 $\pm$ 0.28 & 0.46 &  32.67 &  13.1 & 0.122 & 0.035 &0.18 &  30.527 $\pm$ 0.280 & 32.88 $\pm$ 0.30 &  37.7 $\pm$   5.30\\  
 125 & 4 & 5.28 $\pm$ 0.94 & 0.47 &  32.78 &  11.2 & 0.166 & 0.028 &0.65 &  31.198 $\pm$ 0.380 & 33.24 $\pm$ 0.40 &  44.5 $\pm$   8.30\\  
  58 & 5 & 2.62 $\pm$ 0.24 & 0.92 &  32.72 &  2.85 & 0.184 & 0.096 &0.13 &  31.419 $\pm$ 0.430 &33.73 $\pm$ 0.44 &  55.7 $\pm$    11.0\\   
  68 & 6 & 5.03 $\pm$ 1.49 & 4.13 &  32.72 &  1.22 & 0.144 & 0.396 &0.13 &  31.058 $\pm$ 0.430 & 33.62 $\pm$ 0.47 &  53.1 $\pm$    12.0\\\hline
& & & & & & & & &  &    {\bf 33.28 $\pm$ 0.09} & {\bf 45.3 $\pm$ 2.0}\\
\end{tabular}
\end{center}
\caption[]{\label{revisedcen}The revised SBF distance estimates for the Centaurus cluster galaxies, based on the SBF-measurements from Mieske \& Hilker (\cite{Mieske03b}). The first six columns correspond to Table~4 from Mieske \& Hilker (\cite{Mieske03b}). The revised columns are those of the newly calculated $\Delta {\rm sim}$, $\Delta BG$, $\Delta GC$, and hence $\overline{m}_{\rm I}$, $(m-M)$ and linear distance $d$. We report a typo from Mieske \& Hilker (\cite{Mieske03b}): the colour of CCC 75 is $(V-I)=1.02$ mag, not 1.12 mag.}
\end{table*}
\label{results}
\subsection{Relative distance between Hydra and Centaurus from SBF}
The mean distance of the Hydra cluster is 41.2 $\pm$ 1.4 Mpc, the revised mean distance of the Centaurus cluster is 45.3 $\pm$ 2.0 Mpc. The corresponding mean distance moduli are 33.07 $\pm$ 0.07 mag for Hydra and 33.28 $\pm$ 0.09
mag for Centaurus.
The relative distance in magnitudes then is
$(m-M)_{\rm Cen}-(m-M)_{\rm Hyd}=0.21 \pm 0.11$ mag. The relative distance in Mpc is 
$d({\rm Cen})-d({\rm Hyd})=4.1 \pm 2.4$ Mpc.\\
When excluding distances to galaxies with colours outside the empirically calibrated range $1.0<(V-I)<1.30$ mag, the Hydra and Centaurus distances change by less than 1\%.
The relative distance between the central Centaurus galaxy NGC 4696 and the three Hydra giants NGC 3308, 3309, 3311 is 0.16 $\pm$ 0.19 mag, agreeing very well with the relative distance derived using all galaxies. NGC 4709 was not included in that comparison, as it is redder than the empirically calibrated range and might be located somewhat in front of Centaurus, 
see MH and Sect.~\ref{GCLFcheck} of this paper.
\subsection{Peculiar velocities of Hydra and Centaurus}
\label{H0}
\begin{figure}
\begin{center}
\epsfig{figure=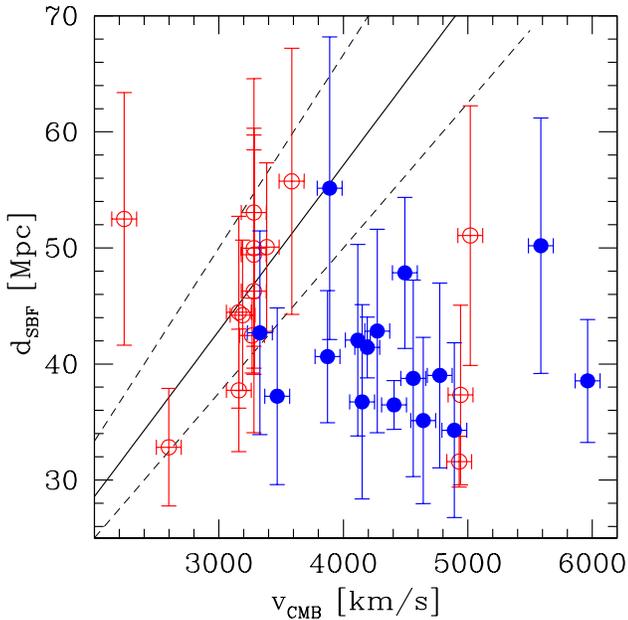,width=8.6cm}
\end{center}
\caption[]{\label{raddm}CMB rest-frame radial velocity 
is plotted vs. SBF distance for 
the Hydra galaxies investigated in this paper (filled circles) and the revised Centaurus galaxies from Mieske \& Hilker (\cite{Mieske03b}) (open circles). 
The solid line gives the Hubble flow 
for $H_{\rm 0}$=70 km~s$^{-1}$ Mpc$^{-1}$. The upper dashed line corresponds to $H_{\rm 0}$=60, the 
lower dashed line to $H_{\rm 0}$=80.}
\end{figure}

To estimate the distortion of the Hubble flow between us and Hydra, we adopt the mean heliocentric radial velocity 3853 $\pm$ 128 km~s$^{-1}$ of the 44
Hydra cluster members within the inner 10 $'$ of the cluster center, based on
the catalog of Christlein \& Zabludoff (\cite{Christ03}). This velocity has to be
corrected for the relative motion of the Sun with respect to the CMB. We
adopt the value derived by Lineweaver et al. (\cite{Linewe96}) from the COBE
dipole CMB anisotropy, who find the Sun moving at 369 km~s$^{-1}$ toward
galactic coordinates $l=264.\degr31$, $b=48.\degr05$. This results in a
CMB dipole velocity component of +336 km~s$^{-1}$ towards the Hydra
cluster and of +281 km~s$^{-1}$ towards the Centaurus cluster. The mean
CMB rest-frame radial velocity of the Hydra cluster then becomes 4190 $\pm$ 128 km~s$^{-1}$.
We assume a cosmological value of $H_{\rm 0}=72 \pm 4$
km~s$^{-1}$ Mpc$^{-1}$ (e.g. Freedman et al  \cite{Freedm01}, Spergel et
al. \cite{Sperge03}, Tegmark et al. \cite{Tegmar04}). Then, our mean distance of 41.2 $\pm$ 1.4 Mpc corresponds to 
a peculiar velocity of 1225 $\pm$ 235 km~s$^{-1}$. \\
In MH the peculiar velocity of Centaurus --
specifically its Cen30 component -- was
calculated with respect to the Sun and not the CMB. We re-calculate the Cen30 peculiar velocity
here in the CMB rest-frame, using the revised Centaurus distances presented in this paper. The mean heliocentric radial velocity of the Cen30 component is 3170
$\pm$ 174 km~s$^{-1}$ (Stein et al. \cite{Stein97}), yielding a CMB velocity of 3450 $\pm$ 174 km~s$^{-1}$. The mean distance of the 8 Cen30 members investigated by us is 45.0 $\pm$ 2.7 Mpc. The cosmological CMB velocity value
corresponding to $H_{\rm 0}=72$ km~s$^{-1}$ Mpc$^{-1}$ is 3240 $\pm$ 240 km~s$^{-1}$. The
peculiar velocity of Cen30 then is 210 $\pm$ 295 km~s$^{-1}$, much smaller than for Hydra
and consistent with an undistorted Hubble flow towards Centaurus. If we include the Cen45 component in our analysis, the peculiar velocity rises slightly to 450 $\pm$ 300 km~s$^{-1}$, still lower than the Hydra velocity by about 800 km~s$^{-1}$ at 2 $\sigma$ significance. Using only the giant galaxies for the peculiar velocity calculation, we obtain 395 $\pm$ 345 km~s$^{-1}$ for NGC 4696 in Cen30 and 1345 $\pm$ 240 km~s$^{-1}$ for Hydra, agreeing very well with the values obtained from the entire sample.\\
In Fig.~\ref{raddm}, a Hubble diagram for Hydra and Centaurus is shown, 
illustrating the higher peculiar velocity for Hydra. Within the uncertainties
of the measured distances, we cannot detect a negative slope between distance and
radial velocity as it would be the case if the more distant galaxies in the sample
would be falling into the Great Attractor from behind. See also the next section for
a discussion on the depth of the Hydra cluster.\\
In Sect.~\ref{discussion}, it will be discussed whether the high peculiar velocity for Hydra together with the lower one found for Centaurus can be explained by the Great Attractor model as proposed by T00.
\subsection{Depth of the Hydra cluster}
\label{depth}
The mean single measurement uncertainty for the Hydra cluster SBF-distances as shown in Table~\ref{resultstab} is 7.4 Mpc, whereas the scatter of these distances around
their mean is 5.7 Mpc. This is consistent with the assumption that the radial extension of the investigated 
portion of the Hydra cluster is small compared to the measurement uncertainty.
To be more specific, we derive an upper limit for the depth of the Hydra cluster applying 
the inequality $\frac{(n-1)^2 (\Delta x)^2}{\chi ^2_{1-\frac{\alpha}{2}}}\le {\sigma}^2 \le \frac{(n-1)^2 (\Delta x)^2}{\chi ^2_{\frac{\alpha}{2}}}$
to obtain the confidence interval at probability $\alpha$ for the real variance
$\sigma ^2$ of a distribution of $n$ measurements
with a measured variance $\Delta x ^2$ (Hackbusch et al.~\cite{Teubne96}). 
From tabulated $\chi^2$ values we find that the Hydra cluster would have to be
radially extended over 3.5 Mpc to both sides in order 
to exclude with 95\% confidence 
a $\delta$-distribution for the distance of our sample galaxies.\\
We therefore derive a formal 
upper limit of $\pm$ 3.5 Mpc radial extension for the Hydra cluster. The Hydra cluster's extension
on the sky is about 3 degrees (Tonry et al. \cite{Tonry00}). At 41 Mpc distance this corresponds to
about 2 Mpc. In case of a spherically symmetric distribution we would therefore expect a radial
distance scatter of the order of $\pm$ 1 Mpc.
I.e. we are only sensitive to a cigar-shape with
the major axis at least 3$-$4 times larger than the minor axis. This is quite
unlikely for a relaxed or nearly relaxed cluster as it is the Hydra cluster
(e.g. Tamura et al.~\cite{Tamura96} and~\cite{Tamura00}, Yamasaki et al.~\cite{Yamasa03}).
\begin{figure*}
\begin{center}
\epsfig{figure=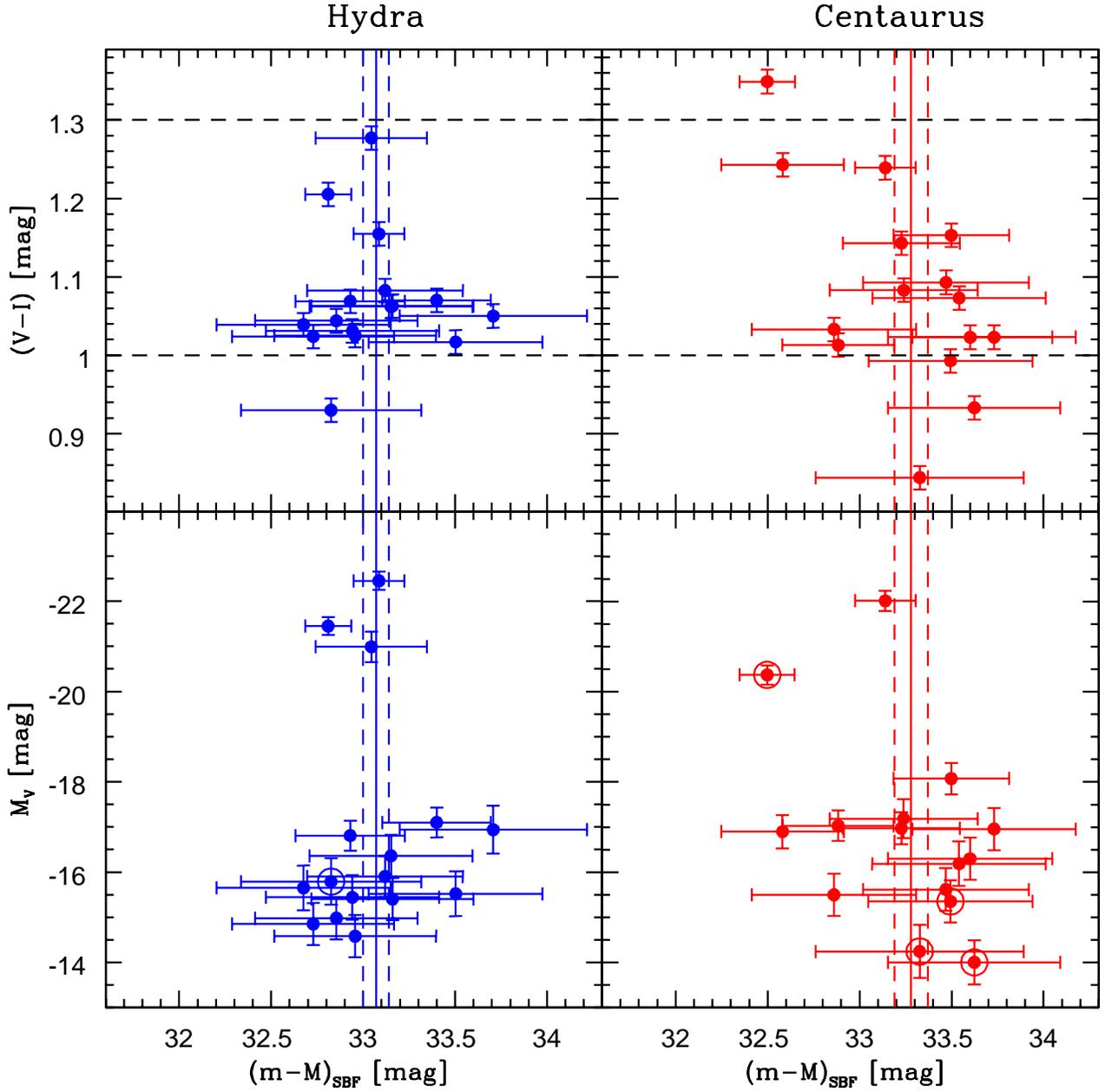,width=17.2cm}
\end{center}
\caption[]{\label{MVVI}{\it Left panels:} Distance modulus $(m-M)_{\rm SBF}$ of the Hydra galaxies
plotted vs. their absolute magnitude $M_V$ (bottom) and colour $(V-I)$ (top). The mean distance
modulus with its 1 $\sigma$ error is indicated by the solid and dashed vertical lines. The dashed horizontal lines in the top panel indicate the colour range of the empirical SBF calibration by Tonry et al.~(\cite{Tonry01}). The data points outside this colour range are marked by open circles in the lower panel.
{\it Right panels:} Plot 
of the same entities as in the left panels for the revised Centaurus cluster distances.}
\end{figure*}
\section{Discussion}
\label{discussion}
\subsection{The Great Attractor acting upon Hydra and Centaurus}
\label{GA}
\begin{figure}[]
\begin{center}
\epsfig{figure=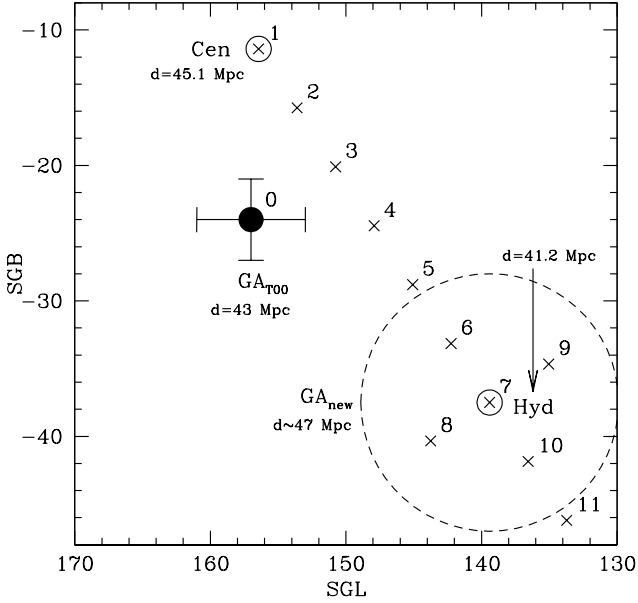,width=8.6cm}
\end{center}
\caption[]{\label{GAmap}Positions in super-galactic
  coordinates 
of the following objects: Great Attractor (GA) according to
Tonry et al. (\cite{Tonry00}, T00 in the following) as filled circle; Hydra cluster as open circle
at $\simeq$ (140,-38); Centaurus cluster as open circle at $\simeq$
(156,-12). The indicated GA distance of 43 Mpc is from T00, the Centaurus distance is based on Mieske \& Hilker
(\cite{Mieske03a}) and revised in this paper. The Hydra distance is from this paper. Crosses and the
attributed numbers indicate assumed projected positions of the GA for which
the expected CMB radial velocities of Hydra and Centaurus are shown
in Fig.~\ref{GAraddif}, as calculated with the flow model by T00. The dashed
circle delimits the region into which the projected position of the GA from 
T00 must be shifted to explain the observed distances and
CMB velocities for Hydra and Centaurus, see Figs.~\ref{GAraddif}~and~\ref{GAplane}. The GA can be accommodated in the entire circle region for a distance of $\simeq$ 47 Mpc.}
\end{figure}

\begin{figure*}[]
\begin{center}
\epsfig{figure=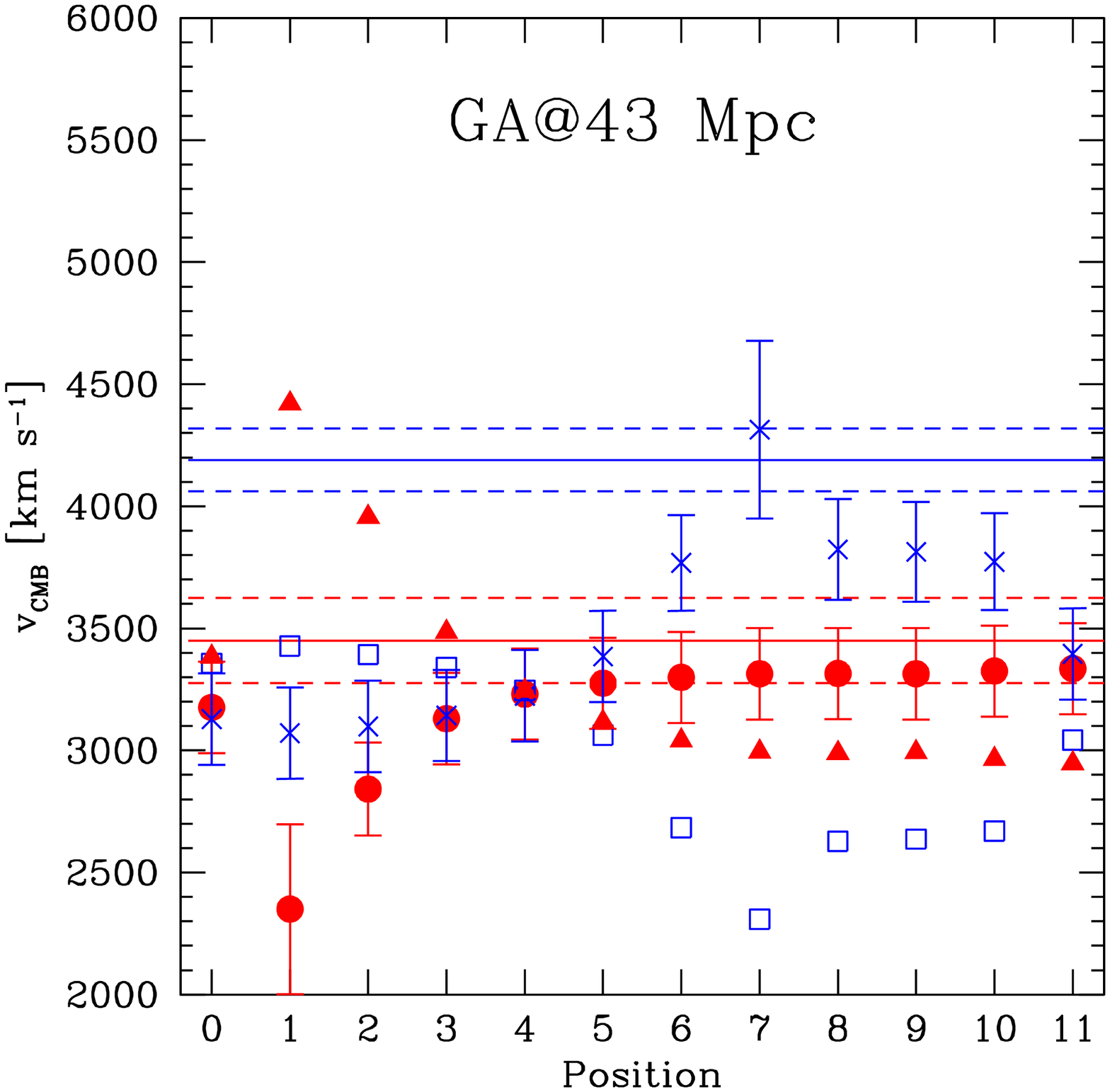,width=4.3cm}
\epsfig{figure=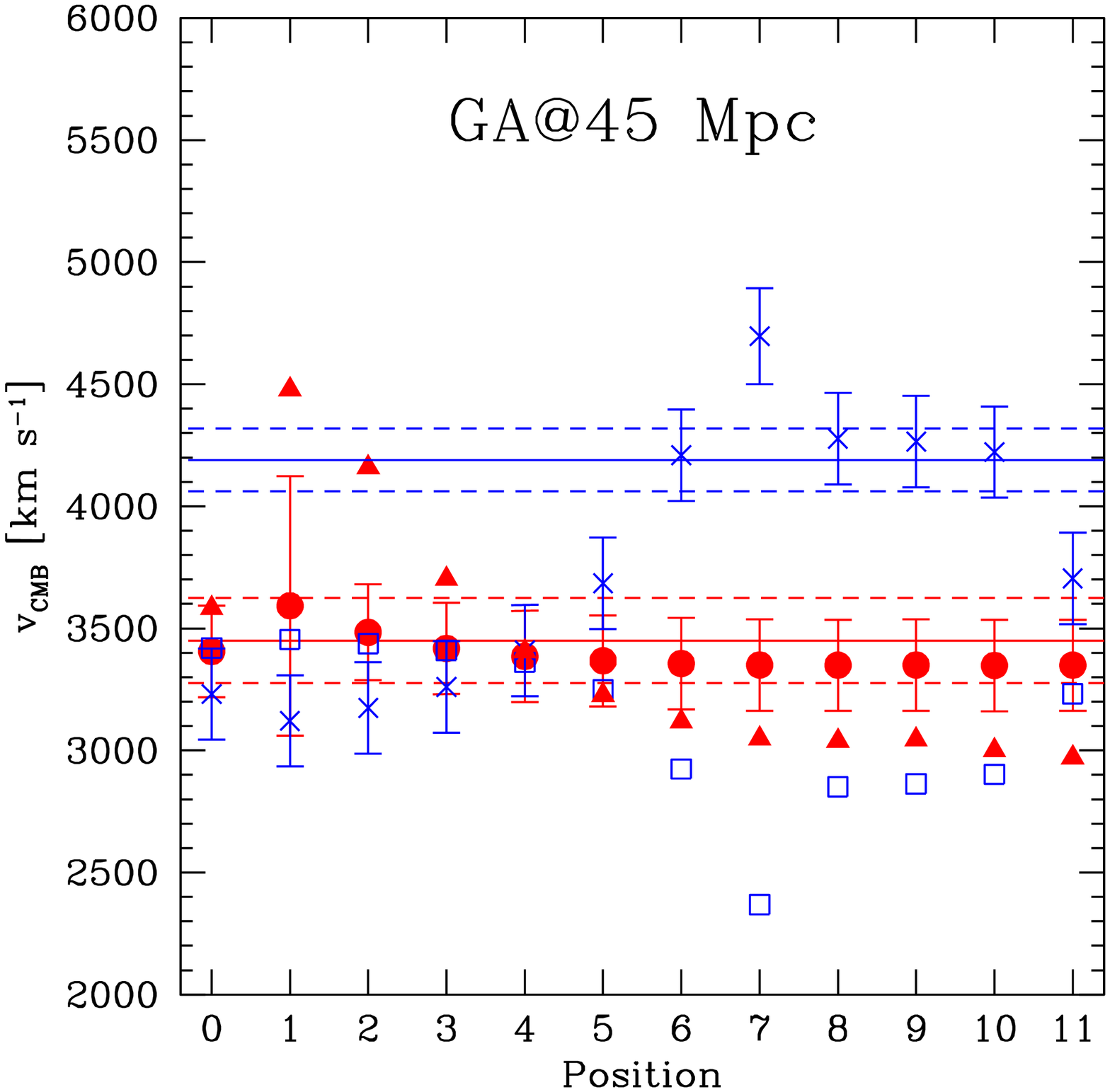,width=4.3cm}
\epsfig{figure=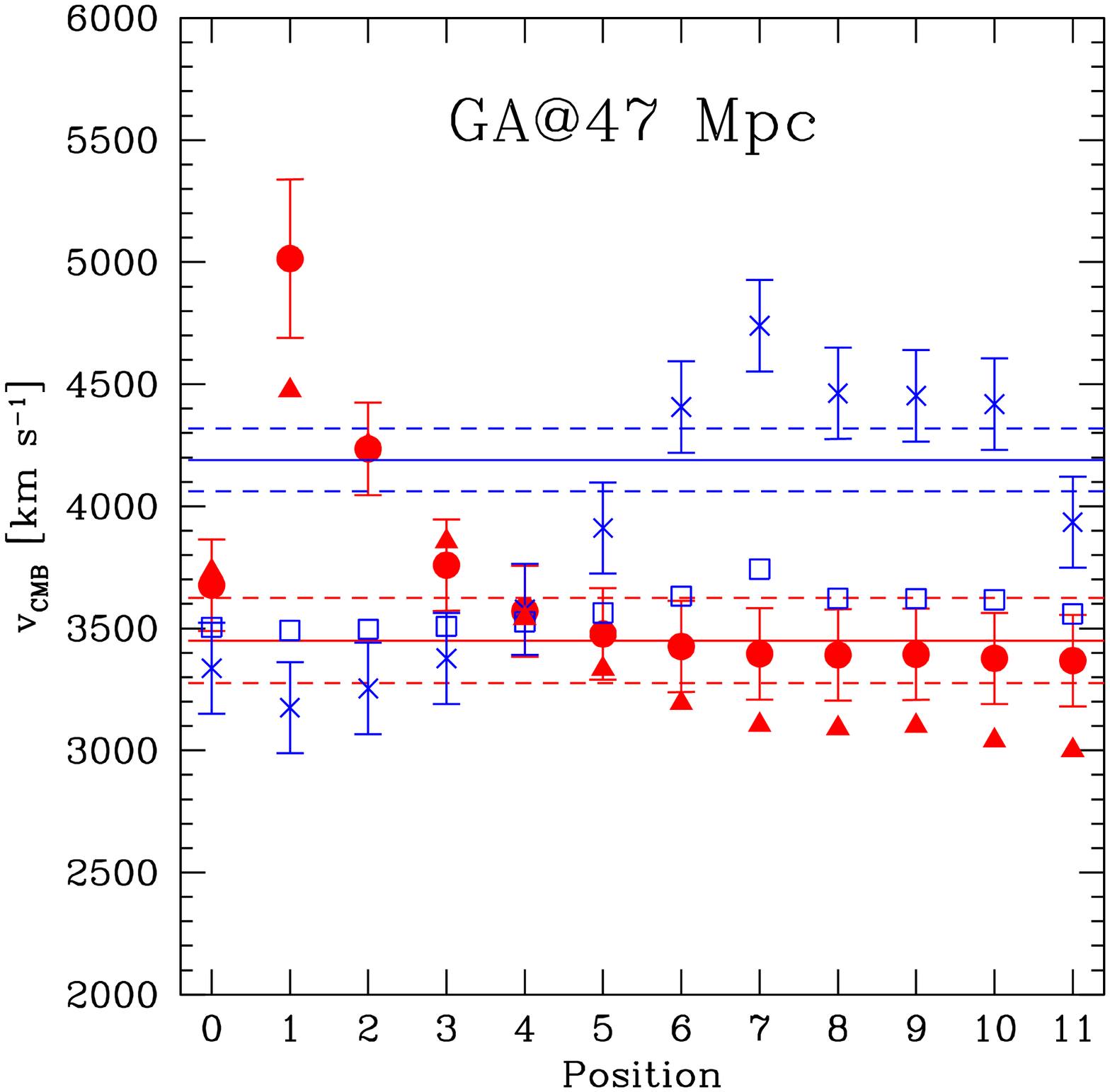,width=4.3cm}
\epsfig{figure=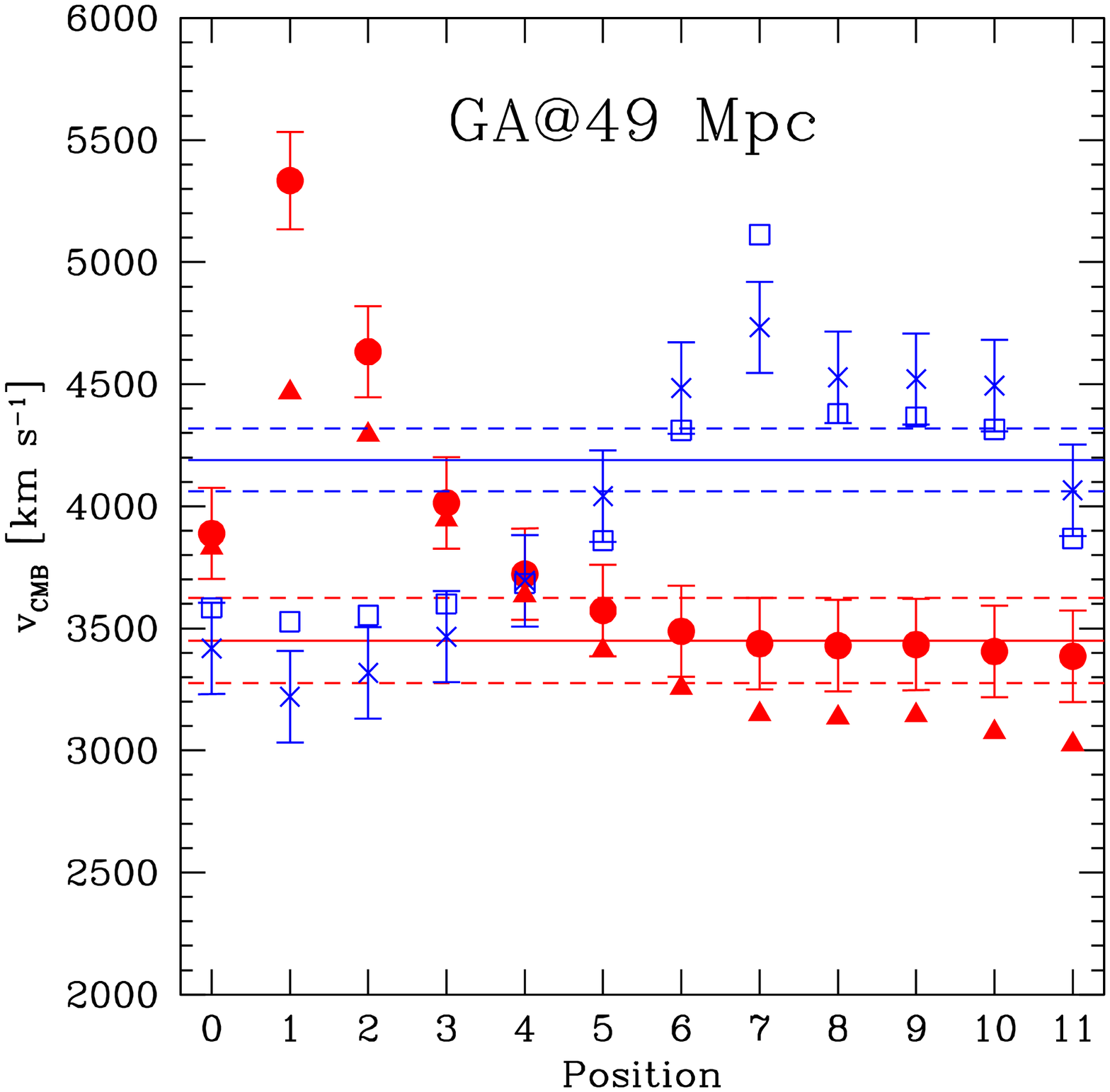,width=4.3cm}
\end{center}
\caption[]{\label{GAraddif}Comparison between expected and measured CMB radial
velocities for four different GA distances of 43, 45, 47 and 49 Mpc. The x-axes indicates the different projected GA positions from Fig.~\ref{GAmap}. 
{\it Filled circles:} expected CMB
radial velocity for the Centaurus cluster based on the flow model by Tonry
et al. (\cite{Tonry00}), using the Centaurus distance of this paper. {\it Crosses:} expected CMB radial velocities for Hydra, using the Hydra distance of this paper. {\it
  Open squares:} expected CMB radial velocities for Hydra assuming a 15\% higher Hydra distance of 47 Mpc. {\it
  Filled triangles:} expected CMB radial velocities for Centaurus assuming a 15\% lower Centaurus distance of 39 Mpc. Error bars have been omitted for these two sets of expected velocities. {\it Upper horizontal solid line:} CMB rest-frame radial velocity of the
Hydra cluster (Christlein \& Zabludoff
\cite{Christ03}). {\it Lower horizontal solid line:} CMB rest-frame radial velocity of the
Centaurus cluster (Stein et al.~\cite{Stein97}). The dashed horizontal lines
indicate the respective error ranges.}
\end{figure*}

\begin{figure}[]
\begin{center}
\epsfig{figure=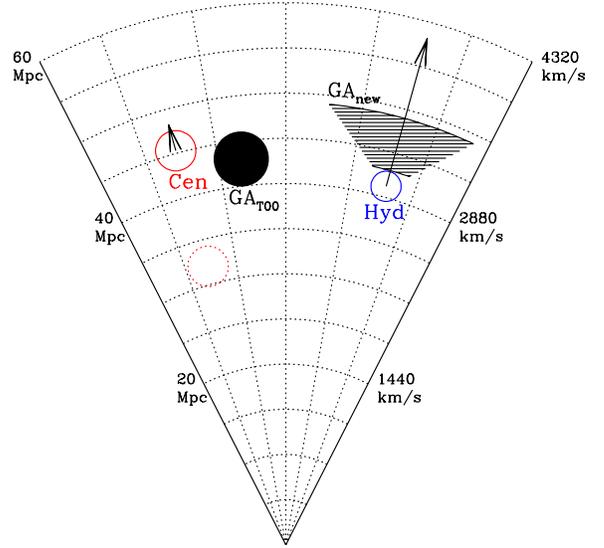,width=8.6cm}
\end{center}\vspace{-1cm}
\caption[]{\label{GAplane}Face-on view towards a plane defined by the positions of Centaurus, Hydra and the Sun, with the Sun being the origin. We assume $H_{\rm 0}=72$ km~s$^{-1}$ Mpc$^{-1}$ to convert distance to CMB redshift. The dotted meridian lines are separated by 10$\degr$. The positions of the Centaurus and Hydra cluster are indicated as circles with solid lines, using the distances derived in this paper. The dotted circle indicates the Cen30 distance derived by Tonry et al.~(\cite{Tonry01}). The symbol sizes corresponds to the distance uncertainties. The position of the Great Attractor from Tonry et al.~(\cite{Tonry00}) is indicated by the filled circle. The peculiar velocities derived in this paper for Hydra and Centaurus are indicated by the arrows. The area within which the Tonry et al. Great Attractor can be accommodated in the light of our new peculiar velocity measurements is indicated by the shaded area, see also Figs.~\ref{GAmap}~and~\ref{GAraddif}.}
\end{figure}

With the Hydra cluster distance derived in this paper and the revised distance value for the Centaurus cluster, some parameters of the Great Attractor (GA) model derived by T00 can be checked. 
The distance to the Cen30 component of
Centaurus derived by us implies a rather low peculiar velocity of 210 $\pm$ 295 km~s$^{-1}$ with respect to an
undisturbed Hubble flow. This has several possible implications for the Great
Attractor model: the mass of the GA is much smaller than the
9$\times$10$^{15}$ M$_{\sun}$ determined by T00; the GA is very massive but the Centaurus cluster falls into the GA almost perpendicular
to the line of sight; or, the Centaurus cluster {\it is} the Great Attractor.\\
The large peculiar velocity of 1225 $\pm$ 235 km~s$^{-1}$ that we derive for the
Hydra cluster and the fact that a possible Hydra cluster infall into Centaurus has only a negligible radial component (see Fig.~\ref{GAplane}) supports the second possibility: a massive GA with its center of mass slightly behind the Hydra-Centaurus plane, in closer projection to Hydra than to Centaurus. Such a location, just like the position determined by T00, implies that the GA is not directly associated with any prominent galaxy cluster, see Figs.~\ref{GAmap}~and~\ref{GAplane}.\\
Before proceeding to a more detailed investigation of this possibility, it is useful to test the null hypothesis that Hydra and Centaurus in reality share a common flow velocity. To do so, we add in quadrature a thermal velocity dispersion component of 200 km~s$^{-1}$ (see T00) to the peculiar velocity errors of both galaxies, and subtract in quadrature the mean error contribution from the uncertainty in $H_0$ from the error of the peculiar velocity difference, as we want to define the uncertainty of the relative peculiar velocity. This yields a relative
peculiar velocity between Hydra and Centaurus of 1015 $\pm$ 440 km~s$^{-1}$. The null hypothesis of a common flow velocity for Hydra and Centaurus is therefore rejected at 2.3 $\sigma$ significance (98\% confidence). \\
 Fig.~\ref{GAmap} shows the projected positions of the Hydra and
Centaurus cluster and the Great Attractor from T00.
The Great Attractor in this flow model
is defined by its 3D-position between Centaurus and Hydra at a distance d=43$\pm$3 Mpc, 
over-density $\delta$, infall exponent
$\gamma$, core radius $r_c$ and cutoff radius $r_{cut}$.  
T00 provide a FORTRAN program which implements
their flow model including the Virgo Attractor, Great Attractor and a
quadruple component. It gives as an output the expected CMB radial velocity
at an input 3D-position. The parameters of all components and the cosmological
parameters $H_{\rm 0}=78$ km~s$^{-1}$ Mpc$^{-1}$ and $\Omega_{\rm m}=0.2$ go into the
calculation. Fig.~\ref{GAraddif} shows these expected CMB radial
velocities at the 3D-positions of Hydra and Centaurus,
for varying projected GA positions and at four different GA distances between 43 and 49 Mpc. Except for the 3D-positions, all other GA parameters were adopted as determined by T00, because we only trace the peculiar velocity field at two 3D positions, which makes it senseless to fit a larger number of parameters. An updated distance set in a much larger area would be needed for that. The four different GA distances assumed are almost within the error range of 43$\pm$3 Mpc given by Tonry for the GA distance. Also, the angular changes of the GA positions in Fig.~\ref{GAmap} are only between Hydra and Centaurus and of the order of 20-30 degrees. They do not significantly affect the gravitational pull of the GA on the Local Group.\\
Figs.~\ref{GAmap} and~\ref{GAraddif} show
that the Hydra and Centaurus distances derived by our group can be explained by Tonry's GA model if the GA is shifted at least 10$\degr$ towards lower super-galactic longitude
and 10$\degr$ towards lower latitude with respect to the original projected GA position determined by T00. This is illustrated also in Fig.~\ref{GAplane}. At an assumed GA distance of 43 Mpc, the GA would have to be directly behind the Hydra cluster in order to exert a sufficient gravitational pull. Already at 45 Mpc GA distance, GA projected positions up to 10$\degr$ away from Hydra are possible. At GA distances of 47 and 49 Mpc --  slightly outside the error range of T00's GA distance --  the expected peculiar velocities rise and start to exclude a GA position directly behind Hydra.
Note that at the original GA position from T00, the expected
radial
velocity for Hydra is below its measured velocity by about 1000 km~s$^{-1}$, ruling out this position.\\
The expected peculiar velocity of Centaurus matches the observed one as soon as the projected GA position is several degrees away from Centaurus, see Figs.~\ref{GAmap} and~\ref{GAraddif}. It also matches the observed one if the GA is identical to the Centaurus cluster (position 1 at 45 Mpc distance). 
However, this latter possibility fails to explain the very large peculiar velocity measured for Hydra. If GA=Centaurus, then it would exert only a very small radial gravitational pull on Hydra, because Hydra's infall vector would be mainly tangential (see Fig.~\ref{GAplane}): Assuming GA=Centaurus and the T00 GA mass of 9$\times$10$^{15}$ M$_{\sun}$ yields a radial infall velocity for Hydra of 130 $\pm$ 70 km~s$^{-1}$, where the uncertainty arises from the error in relative distance between Hydra and Centaurus.
The assumption that in addition to a common flow of both clusters -- which is rejected at 98\% confidence -- there is a 200 km~s$^{-1}$ Hydra infall into Centaurus (=GA) is still rejected at 94\% confidence. In this context one important notion is that the mass of the Centaurus cluster derived from X-ray temperature maps ($\simeq 3 \times 10^{14} M_*$, see for example Reiprich \& B\"ohringer~\cite{Reipri02}) is about a factor of 30 lower than the 10$^{16}$ M$_{*}$ estimated for the GA in T00. This a priori makes the Centaurus cluster an improbable GA candidate. We note that only in the case of an unrealistically massive GA at the Centaurus position with 5 $\times$10$^{16}$ M$_{\sun}$  - which is totally inconsistent with the amount of peculiar velocities observed in the nearby universe -, the confidence level of rejecting a common flow for Hydra and Centaurus drops to 1$\sigma$. A radial infall velocity for Hydra of 500 $\pm$ 270 km~s$^{-1}$ into the GA would be expected in this case.\\
From the above it is clear that a simultaneous check of more than one parameter (mass, distance, infall parameter, etc.) of the Tonry GA can only be performed with an updated distance set extending over a much larger area  and distance range than sampled by us. Note, however, that only a change in projected position is {\it necessary} to explain the peculiar velocities of Hydra and Centaurus observed by us. To nevertheless get an idea on the order of the degeneracies involved when using only two 3D positions, we estimate the GA mass-distance degeneracy for projected GA positions close to Hydra (positions 6 to 11 in Fig.~\ref{GAraddif}): at a GA distance of 47 Mpc, a smaller GA mass by a factor of two would still not underpredict the peculiar velocity for Hydra, while at a GA distance of 43 Mpc, a larger GA mass by a factor of two would still not overpredict the Hydra peculiar velocities.\\
The shift of the projected GA position that is required to obtain
consistency with our peculiar velocity measurements is almost three times
larger than its 3 $\sigma$ confidence range from T00. 
Note, however, that T00 did not include any
Hydra cluster galaxy into their model calculations. 
This stresses that a recalculation of the flow model might become
necessary in the light of increasingly more high-quality SBF data being
published.\vspace{0.2cm}\\
Summarising, the two SBF-distances for Hydra and Centaurus derived by our group
rule out a common flow velocity for these two clusters at 98\% confidence. Within the scenario of a Great Attractor (GA) somewhere in the Hydra-Centaurus region, our results are inconsistent at 94\% confidence with a picture where Centaurus is identical to the GA, both Centaurus and Hydra share a common flow and Hydra has an additional infall into the GA. Our results are consistent with a shift in projected GA position by at least 15$\degr$ towards the Hydra cluster compared to the position determined by Tonry et al.~(\cite{Tonry00}), see Figs.~\ref{GAmap}~and~\ref{GAplane}. A change in the GA mass or distance within the error ranges of Tonry's flow model is not required.\\
Our results increase the angular distance between the Hydra-Centaurus GA and the overdense region in the Zone of Avoidance (ZOA), especially the Norma cluster, to about 50$\degr$. As the Norma cluster has also been proposed as the possible center of a ``Great Attractor'' region (Woudt et al.~\cite{Woudt03}, Kolatt et al.~\cite{Kolatt95}), this large angular separation supports the idea that the SBF-GA and the ZOA-GA might be different substructures within a generally overdense region of the universe.
\subsection{Some other ideas}
The main result of the previous section is that the Great Attractor is probably not identical to neither Hydra nor Centaurus. At face value this implies the existence of a super-massive invisible dark halo, in contradiction to the widely accepted paradigm that light traces matter reasonably well on large scales. We note that the same problem also holds for the position of the Great Attractor from the study of Tonry et al.~(\cite{Tonry00}). However, one must keep in mind that this scenario is a rather simplified picture of reality. The mass distribution in the universe is continuous and the formal position of the Great Attractor should be considered the center of mass of the entire -- continuous -- mass distribution in that particular region. Nevertheless, the absence of a prominent galaxy cluster in the background of Hydra and Centaurus is puzzling with respect to our results.\\
One possibility to explain a high Hydra cluster peculiar velocity without the need for a ``dark'' Great Attractor would be infall along a filamentary structure. However, this scenario would require that our sample of galaxies in the direction and within the redshift range of Hydra be biased towards those with lower distances. In Sect.~\ref{syseff} we show that this is not the case. Furthermore, Hydra is the prototype of a relaxed cluster (Tamura et al.~\cite{Tamura00}, Yamasaki et al.~\cite{Yamasa03}) and is therefore unlikely to have a radial extension significantly longer than the tangential one.\\
As shown in Sect.~\ref{GA}, one improbable (2\%) -- but not impossible -- scenario is that both Hydra and Centaurus take part in a common bulk flow and that the large measured difference in peculiar velocity is caused by the thermal velocity field and our measurement errors. This scenario becomes more likely if in addition to the random peculiar velocity field we assume that both clusters were formed in the same primordial dark matter halo which carried away a net angular momentum which then was distributed among the Hydra and Centaurus sub-clumps. A crude estimate of additional peculiar velocity introduced into the Hydra-Centaurus system can then be obtained by assuming pure Keplerian rotation of both clusters around their center of mass. Assuming the X-ray mass estimates for Hydra and Centaurus by Ettori et al.~(\cite{Ettori02}) and Reiprich \& B\"ohringer (\cite{Reipri02}) yields a total mass of $\simeq 6\times 10^{14} M_{\sun}$. The projected distance between both clusters is about 20 Mpc. The rotational velocity on a circular orbit of radius 10 Mpc with a central mass of $6 \times 10^{14} M_{\sun}$ is about 510 km~s$^{-1}$. This is of the order of the peculiar velocity difference measured by us. However, the Hydra-Centaurus system is much too large to assume that both clusters are gravitationally bound to each other: the time for one revolution on a circular orbit would be around 90 Gyrs. Furthermore, one crossing time is about 30 Gyrs at 500 km~s$^{-1}$ velocity. Due to these large distances and long time-scales, it seems inadequate to consider Hydra and Centaurus as correlated sub-clumps of one main primordial dark matter halo. Therefore, adding in quadrature a thermal peculiar velocity component to the peculiar velocity errors of both Hydra and Centaurus (see Sect.~\ref{GA}) seems to be the proper way to account for primordial inhomogeneities in the matter and momentum distribution.
\subsection{Comparison with literature distances}
\label{complit}
As our distance for the Hydra cluster implies a very large peculiar
velocity, we compare both our Hydra and Centaurus distance with values obtained by other authors.\\
\subsubsection{Centaurus}
\label{centaurus}
Compared to our revised Centaurus distance of 33.28 $\pm$ 0.09 mag, 
Tonry et al. (\cite{Tonry01}) obtained a significantly 
smaller distance by more than 0.5 mag for their Centaurus
sample, partially attributed to a Malmquist-like bias (see discussion in MH). 
For the two galaxies NGC 4696 and NGC 4709 common to both surveys, our mean
distance is 32.825 mag, less than 0.1 mag higher than the mean Tonry distance. As extensively discussed in MH and confirmed in this paper from both SBF and GCLF distances, NGC 4709 appears to be located in front of the Centaurus cluster, which is why our mean distance of NGC 4696 and NGC 4709 is significantly lower than the mean cluster distance.\\
Pahre et al. (\cite{Pahre99}) present
an SBF distance measurement for a Centaurus cluster galaxy which was not mentioned
in MH: for the giant elliptical NGC 4373, 
Pahre et al. obtain
$(m-M)=$ 32.99 $\pm$ 0.11 mag from WFPC2 images, slightly lower than both our distance to NGC 4696 and the entire cluster.\\
Another distance estimate for the Centaurus cluster comes from the Fundamental Plane (FP) 
analysis presented within the SMAC survey 
(e.g. Smith et al.~\cite{Smith00}, \cite{Smith01}, \cite{Smith04} and 
Hudson et al.~\cite{Hudson04}), which includes peculiar velocity measurements 
for both the Hydra and Centaurus cluster. 
In the framework of this FP analysis, the difference between $cz_{FP}$ from the FP analysis and the measured 
$cz_{CMB}$ 
of the cluster yields its peculiar radial velocity with respect to the Hubble flow.
For comparison of the
SMAC results with our metric distances, we adopt
$H_{\rm 0}=72 \pm 4$ km~s$^{-1}$ Mpc$^{-1}$. For Centaurus, the revised mean SBF distance is
45.3 $\pm$ 2.0 Mpc, 
while the SMAC result (Smith et al.~\cite{Smith04}, priv. comm.) is 
$cz_{FP}= 3019 \pm 158$ km~s$^{-1}$, corresponding to 41.9 $\pm$ 3.2 Mpc for 
$H_{\rm 0}=72 \pm 4$ km~s$^{-1}$ Mpc$^{-1}$. Both values agree very well.\\
For the Centaurus
cluster we can therefore state that the SMAC FP-distance is consistent with our SBF-distance, implying a small peculiar velocity for Centaurus. Literature SBF-distances are slightly below our value by about 15\%.
\begin{figure}
\begin{center}
\epsfig{figure=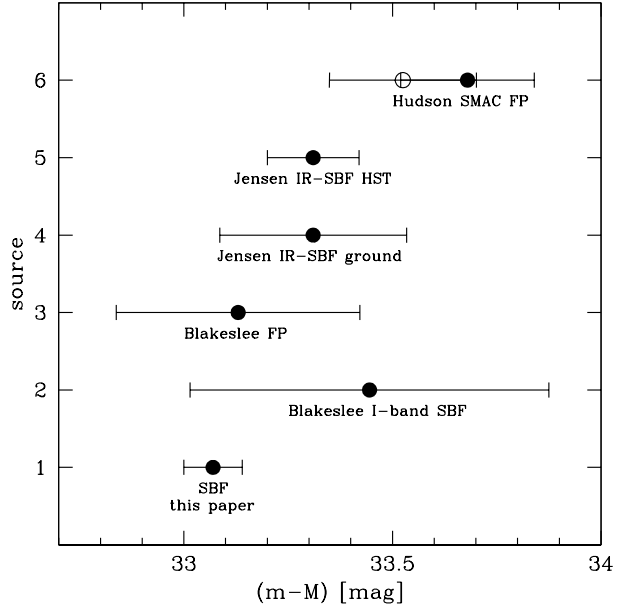,width=8.6cm}
\end{center}
\caption[]{\label{distdif}Literature comparison of distances to the Hydra cluster.
Numbers refer to the following sources: 1 SBF distance from this paper; 2 mean I-band SBF distance to
NGC 3309 and NGC 3311 from Blakeslee et al. (\cite{Blakes02}, and priv. comm.); 3 mean
FP distance to NGC 3309 and NGC 3311 from Blakeslee et al. (\cite{Blakes02}); 4 mean K-band
SBF distance to NGC 3309 and NGC 3311 from Jensen et al. (\cite{Jensen99}); 5 NICMOS F160W
SBF-distance to NGC 3309 from Jensen et al. (\cite{Jensen01}) using the revised calibration by Jensen et al. (\cite{Jensen03}); 6 FP distance to
Hydra cluster from Hudson et al. (\cite{Hudson04}), open symbol refers to the distance
when including the Mg$_2$ term in the FP-relation fit.}
\end{figure}
\subsubsection{Hydra}
There are four recent publications that present
distances to either the entire Hydra cluster
or single cluster galaxies, see also Fig.~\ref{distdif}:\\
1. Blakeslee et al. (\cite{Blakes02}), who perform a comparison between SBF and Fundamental
Plane distance for early type galaxies, including the Hydra giant ellipticals 
NGC 3309 and NGC 3311. 
 The SBF distances are 33.66 $\pm $ 0.68 mag for
NGC 3309 and  33.23 $\pm$ 0.52 mag for NGC 3311 (Blakeslee, private communications). This corresponds to a mean distance
of 49.1 $\pm$ 10 Mpc, which is larger than but still consistent with our mean Hydra distance.
The FP distances are 33.33 $\pm$ 0.41 mag
for NGC 3309 and 32.93 $\pm$ 0.41 mag for NGC 3311. This corresponds to a
mean distance of 42.4 $\pm$ 5.8 Mpc and is in good agreement with our
results.\\
2. Another distance estimate for the Hydra cluster comes from the SMAC survey.
For the Hydra cluster, Smith et al. (\cite{Smith04} and priv. comm.) 
obtain $cz_{FP}= 3919 \pm 206$ km~s$^{-1}$.
This corresponds to a distance of 54.4 $\pm$ 4.2 Mpc when assuming 
$H_{\rm 0}=72 \pm 4$ km~s$^{-1}$ Mpc$^{-1}$. 
This is inconsistent with our distance of 41.2 $\pm$ 1.4 Mpc at the 3.0 $\sigma$ level and
corresponds to a difference of 0.60 $\pm$ 0.18 mag in distance modulus.\\
3. Jensen et al. (\cite{Jensen99}), who measure IR-SBF
distances for NGC 3309 and NGC 3311 in the $K$'-band using the Hawaii 2.24m telescope.
They obtain identical distances to NGC 3309 and NGC 3311 of 46 $\pm$ 5 Mpc, using
$\overline{M}_{\rm K'}=-5.61 \pm 0.12$ 
as derived in Jensen et al. (\cite{Jensen98}). This is higher than our Hydra cluster distance by about 10\%, but still marginally consistent. It is about 20\% higher than our mean distance to NGC 3309 and NGC 3311. A possible reason
 for the higher distance may be that the SBF signal measured by Jensen
et al. was in some parts dominated by background spatial variance, whose removal
was estimated to introduce errors of up to 0.4 mag in the finally adopted value
for $\overline{m}_{\rm K'}$. Also, the more recent calibration of 
$\overline{M}_{\rm K'}$ by Liu et al. 
(\cite{Liu02})
shows an intrinsic scatter of about 0.25 mag in $\overline{M}_{\rm K'}$ 
for the galaxies observed, 
about twice as large as assumed by Jensen et al. (\cite{Jensen99}).\\
4. Jensen et al. (\cite{Jensen01}), who obtain NICMOS F160W SBF distances of 16 
distant galaxies with $cz<10000$ km~s$^{-1}$, among them NGC 3309 as the only Hydra cluster
member. For this galaxy they measure an apparent fluctuation magnitude of
$\overline{m}_{\rm F160W}=28.86 \pm 0.07$ mag. 
To convert this into $\overline{M}_{\rm F160W}$, we use the revised calibration of 
Jensen et al. (\cite{Jensen03}). For the colour range $1.05<(V-I)<1.24$ mag, they derive
\begin{equation}
\label{sbfir}
\overline{M}_{\rm F160W}=(-4.86\pm 0.03)+(5.1\pm 0.5) \times [(V-I)_0 - 1.16]
\end{equation}
This is equation~(1) of their paper. In Jensen et al. (\cite{Jensen01}), $(V-I)_0$ of NGC 3309
is estimated from $(B-R)$ colours by Postman \& Lauer (\cite{Postma95}) to be 1.28 mag.
This is slightly redder than the value of 1.21 mag determined by us directly via $VI$ photometry in this paper. From our
photometry we detect a slight colour gradient with redder colour towards the center of 
NGC 3309.
The region sampled by Jensen et al. (\cite{Jensen01}) for SBF measurement is a ring region
with $2.4 \arcsec <r<4.8 \arcsec$, which is closer to the center than even our innermost ring. 
Our photometry
indicates a colour of $(V-I)\simeq$ 1.23-1.24 mag in this area. Adopting 1.24 $\pm$ 0.02 mag 
and plugging
this value into equation (\ref{sbfir}) results in $\overline{M}_{\rm F160W}=-4.45 \pm 0.05$ mag
for NGC 3309. This yields a distance estimate of $(m-M)=33.31 \pm 0.11$ mag.
This is higher than our mean 
distance for Hydra by 0.24 mag (12\%) and higher than the distance to NGC 3309 by about 0.5 mag.
\subsubsection{Discussion of the Hydra distance}
Although for the Hydra cluster there is a substantial disagreement between our SBF distances
and the SMAC FP distances, both methods yield practically identical results for the Centaurus
cluster (see Sect.~\ref{centaurus}).
The simultaneous agreement for Centaurus and strong 
disagreement for Hydra is remarkable, given that our Centaurus
cluster data were obtained with the same instrument and filters as the Hydra data and the
SBF measurement procedure was identical for both data sets. In that context
it is very interesting to note that the SMAC value of $cz_{FP}$ for Hydra 
implies a very small peculiar velocity for Hydra, almost consistent with zero.
Hudson et al. (\cite{Hudson04}) show that for the determination of the bulk
flow velocity amplitude, the Hydra cluster is the most significant outlier of all
56 investigated clusters. Excluding this single cluster from the bulk-flow analysis
increases the bulk flow velocity by more than 100 km~s$^{-1}$ up to almost 800 km~s$^{-1}$.\\
The question arises whether the Hydra cluster exhibits peculiarities in the 
stellar populations of its galaxies
that may have a biasing effect on IR-SBF distances and/or FP distances.\\
Already Jensen et al. (\cite{Jensen03}) have noted that IR-SBF measurements 
are more sensitive to age-metallicity
effects than I-band SBF. The spread of $\overline{M}_{\rm F160W}$ at
a given colour can be up to 0.2 mag according to their calibrations. 
Stellar population models (e.g. Liu et al.~\cite{Liu00} and ~\cite{Liu02}, Blakeslee
et al. \cite{Blakes01}) 
indicate that at red colours ($(V-I)\ge$1.2), the $\overline{M}_{\rm F160W}$
values used for the calibration correspond to very old galaxies with age ~15 Gyr.
This means that there is little room for fainter $\overline{M}_{\rm F160W}$ at red colours.
Indeed, Jensen et al. (\cite{Jensen03}) attribute the spread in $\overline{M}_{\rm F160W}$
largely to the distance uncertainty from the
I-band SBF distances used to calibrate the $\overline{M}_{\rm F160W}$ values.
The age-metallicity spread could therefore contribute up to
0.2 mag to the 0.2-0.5 mag difference between
our I-band SBF distances and the IR-SBF-distances for NGC 3309 and 3311.\\
Also Hudson et al. (\cite{Hudson04}) remark that if 
there are systematic age-metallicity differences
between different clusters, this could significantly bias the derived FP 
distance values. As an increase of metallicity tends to fainten a galaxy's surface brightness,
a metal rich outlier would imply a too large distance. A bias like
that can be partially corrected for by including the $Mg_2$ index in the 
inverse FP analysis. When doing so, Hudson et al. (\cite{Hudson04})
 get $cz_{FP}= 3648 \pm 229$ km~s$^{-1}$ for the Hydra cluster.
This corresponds
to a distance of 50.7 $\pm$ 4.3 Mpc, about 4 Mpc less than without the $Mg_2$ index.
The disagreement with our result decreases to 0.45 $\pm$ 0.18
mag, still significant at the 2.1 $\sigma$ level.\vspace{0.2cm}\\
Summarising this subsection, the distance estimates derived for the
Hydra cluster in the last 5-7 years by other authors are between 0.1 and 0.6 mag higher than our result, with a mean of 33.37 mag.
Fig.~\ref{distdif} illustrates this.
Possible systematic effects biasing the results of these authors towards higher distances
are of the order of 0.2-0.4 mag.
\subsection{Systematic effects in our data?}
Having found that our Hydra cluster distance is at the low end and the Centaurus distance at the high end of recent distance estimates by other authors, we discuss in this section
whether and to what extent any systematic biases in our data set may influence
our measurements.\\
\label{syseff}
\begin{figure}
\begin{center}
\epsfig{figure=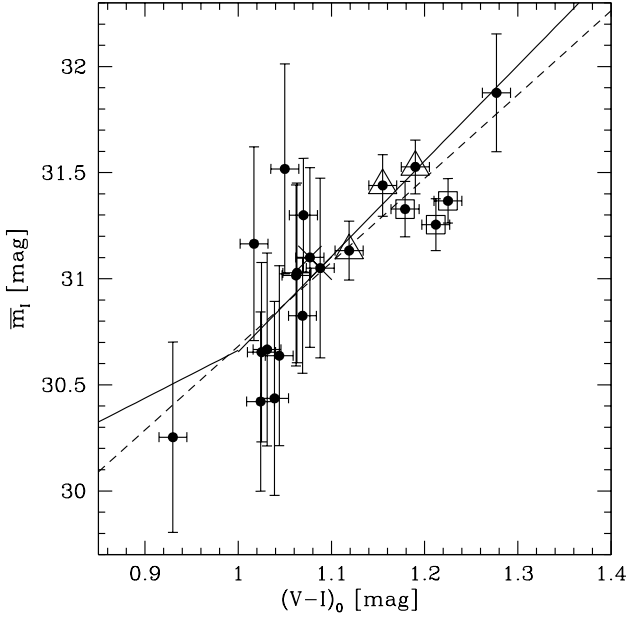,width=8.6cm}
\end{center}
\caption[]{\label{vimi}Deredenned colour $(V-I)_0$ plotted vs. apparent fluctuation
magnitude $\overline{m}_I$ for the investigated Hydra galaxies. Large squares indicate the results obtained
in the three different rings for NGC 3309; large triangles refer to NGC 3311; the two
crosses for the galaxy with $(V-I)_0\simeq 1.09$ mag refer to galaxy No. 252 in field 3,
which was also imaged and investigated in the adjacent field 2.
The solid line indicates the calibration relation adopted
between $\overline{M}_I$ and $(V-I)_0$, shifted to $(m-M)=33.07$ mag, see Sect.~\ref{sbfmeas}.
The dashed line is a linear fit to the
data points, allowing both the zero-point and slope to vary. The fitted slope varies by 0.9 $\sigma$ from the slope adopted in the calibration.}
\end{figure}
\subsubsection{Selection bias?}
One possible effect inherent in our data might be a selection bias towards brighter SBF magnitudes 
at the faint magnitude limit of our survey. Galaxies whose observational uncertainties or
stellar populations place them at seemingly low distances will have a higher S/N in the
SBF measurement than galaxies of the same apparent brightness who fall behind the observed cluster 
due to observational errors. Fig.~\ref{MVVI} shows the absolute brightness $M_V$ of the Hydra galaxies plotted vs.
$(m-M)_{SBF}$. Applying a linear fit to the data points results in a non-zero slope at 0.4 $\sigma$
significance. When rejecting the three faintest data points, the significance changes to 0.6 $\sigma$. 
The mean distance of the three giants is 39.5 $\pm$ 1.5 Mpc, while for the 13 
dwarfs it is 41.6 $\pm$ 1.7 Mpc, i.e. a difference of 5 $\pm$ 6\%. Rejecting the three faintest data points, the mean distance rises by about 2\% to 42.1 $\pm$ 2.4 Mpc. Excluding the two brightest dwarfs lowers the mean distance by about 3\% to 39.8 $\pm$ 1.1 Mpc. All the mean distances for the sub-samples mentioned are within the error ranges of the entire sample.
As noted in Sect.~\ref{datareduction}, there is only one faint early-type Hydra cluster member from the spectroscopic sample of Christlein \& Zabludoff~(\cite{Christ03}) for which
SBF could not be measured due to an undetectable SBF signal. 
This galaxy would have $M_V\simeq -13.5$ mag at the Hydra cluster distance, about 1 mag fainter than
the magnitude regime investigated here.\\
Our results do therefore not imply a significant selection bias for the Hydra galaxies, neither towards too bright nor too faint SBF magnitudes. The same holds for Centaurus, where the distance to NGC 4696 and the dwarfs differs by 9\% $\pm$ 9\% (for NGC 4709 see discussion in Sect.~\ref{GCLFcheck}), and the mean distance is lowered by only 3\% when rejecting the three faintest galaxies.\\
A final check is to calculate the two measures of the observation quality
for SBF-data '$Q$' and '$PD$', defined in T00. SBF-Measurements with $Q<0$ or $PD>2.7$ are considered as potentially affected by low quality. $PD$, which is the product of the seeing-FWHM in arcsec and the radial velocity in units of 1000 km~s$^{-1}$, is about 2.0 for Hydra and about 1.5 for Centaurus (using the undisturbed Hubble flow radial velocity that corresponds to our distance value). The parameter '$Q$', which is the logarithm to the basis of 2 of the ratio of the detected electron number corresponding to $\overline{m}_I$ and PD$^2$, is below 0 for seven Centaurus cluster galaxies. The mean SBF-distance of these seven galaxies is 46.4 $\pm$ 3 Mpc, in perfect agreement with the overall mean. In the Hydra sample, five galaxies have $Q<0$. Their mean distance is 44.1 $\pm$ 3.4 Mpc, also consistent with the overall mean.
\subsubsection{Calibration bias?}
Another possible systematic effect could be that the calibration between
$\overline{M}_I$ and $(V-I)_0$ adopted in this paper might not be valid for
the entire investigated colour range. 
There might be a change in the integrated age/metallicity combination
 between giants and dwarfs that could result in a 
somewhat stronger SBF and hence underestimated distance with respect to the
original calibration. 
In Mieske et al. (\cite{Mieske03a}) and MH, stellar population
models by Worthey (\cite{Worthe94}) are used to show that for colours around
$(V-I)\simeq 1.0$,
age differences of several Gyrs at mean ages of about 10 Gyrs -- with the
corresponding metallicity change -- cause a scatter in
$\overline{M}_{\rm I}$ of the order of 0.2-0.3 mag at constant colour.
When excluding all galaxies with
$(V-I)_0<1.05$ mag from the Hydra sample investigated in this paper, the resulting mean distance is 43.1 $\pm$ 1.9 mag, about 5\% 
or 0.1 mag higher than for the entire sample. This difference is consistent
with the scatter expected from the Worthey models. Doing the same for Centaurus lowers its mean distance by 5\%.\\
Note that a linear fit to the $(m-M)_{SBF}$ - $(V-I)_0$ Hydra data points in Fig.~\ref{MVVI} yields no measurable correlation. Also, excluding the SBF-distances for galaxies with colours outside the empirically calibrated range $1.0<(V-I)<1.3$ mag does not change the mean distance, neither for Centaurus nor for Hydra.
Furthermore, Fig.~\ref{vimi} shows that the slope in the $\overline{m}_I$ $-$ $(V-I)_0$ plane for the Hydra
sample is consistent with the value of 4.5 used for the calibration. The same holds for the Centaurus data in the colour range $1.0<(V-I)<1.3$.\\
Summarising, we do not find indications for significant stellar population effects in our data, neither for Hydra nor for Centaurus. 
Nevertheless, we cannot exclude a bias 
of up to 0.1 mag towards too low distance for Hydra and towards too high distance for Centaurus. 
\subsubsection{Malmquist bias?}
Blakeslee et al. (\cite{Blakes02}) note that both 
their SBF survey as the FP distances from the SMAC survey
are subject to the so-called 'Malmquist-bias' (Malmquist \cite{Malmqu20}). In this context,
'Malmquist-bias' refers to the fact that the expectation value for the true
distance $r$ of a galaxy tends to be higher than its measured distance $d$, the higher the observational
errors are and the closer it lies to the survey limit 
(Blakeslee et al. \cite{Blakes02}, Strauss \& Willick \cite{Straus95}, 
Lynden-Bell et al. \cite{Lynden88}). The Malmquist bias is most severe for distance estimates
to field galaxies that are not bound in a cluster.
To investigate whether it might affect our distance estimates to the Hydra cluster
galaxies,
we quote the formula given by Blakeslee et al. 
(\cite{Blakes02}) for the calculation of $r$ at a measured
value of $d$, based
on the notation of Strauss \& Willick (\cite{Straus95}):\\
\begin{equation}
E(r|d)=\int_{0}^{\infty}\frac{r^3 n(r) exp(\frac{[ln(r/d)]^2}{2 \Delta ^2})}{r^2 n(r) exp(\frac{[ln(r/d)]^2}{2 \Delta ^2})}\mathrm{d}r
\label{malmquist}
\end{equation}
$n(r)$ is the real-space density distribution of galaxies in the direction of the given sample galaxy, 
and 
$\Delta$ is the fractional error in the distance measurement of this galaxy.\\
For the radial density distribution of the Hydra galaxies we adopt:
\begin{equation}
n(r)=exp(\frac{-(d-d_0)^2}{2 \sigma ^2}) + b.
\label{denschoice}
\end{equation}
The first term is the density distribution within the Hydra cluster,
parametrised by the mean distance $d_0$ and width $\sigma$. The constant
$b$ is included to allow for a uniformly distributed population of back- and foreground
galaxies. Plugging equation~(\ref{denschoice}) into equation~(\ref{malmquist})
allows one to iteratively determine the magnitude of the Malmquist bias.
Evaluating equation~(\ref{malmquist}) with $\sigma=$ 1 Mpc (expected for the
Hydra cluster galaxies if they are distributed in a spherically symmetric
manner, see Sect.~\ref{depth}) yields a vanishing Malmquist bias for $b=0$
and only 2\% bias
for 25\% background contamination. This is negligible compared to the
error of Hydra's mean distance. 
The negligible Malmquist bias for $\sigma=$ 1 Mpc still
holds if we multiply our distance errors by 1.5, matching the errors of the SBF-distances
to NGC 3309 and NGC 3311 by Blakeslee et al. (\cite{Blakes02}).\\
Therefore, it is likely that neither in our
investigation nor in that by Blakeslee et al., a Malmquist bias artificially decreases
the distance values to the Hydra galaxies.
\subsubsection{SBF-distance vs. GCLF distance, NGC 4709}
\label{GCLFcheck}
An interesting consistency check of our SBF-distances is to compare them with the GCLF distances derived for the four giant galaxies NGC 3309, NGC 3311 (Hydra), NGC 4696 and NGC 4709 (Centaurus), see Table~\ref{resultsgcs}. For NGC 4696 and 4709, GCLF- and SBF-distances agree to within 0.2 mag. The relative distance between both galaxies from their SBF-measurements is also recovered well by the GCLF-distances. The mean difference $(m-M)_{SBF}- (m-M)_{GCLF}$ for all four giants is 0.31 $\pm$ 0.54 mag. The large error in that value is because the GCLF-distances to NGC 3309 and 3311 are only very poorly constrained, with formal uncertainties around 0.6-0.7 mag. This inhibits a thorough comparison of GCLF- and SBF-distances for the Hydra galaxies. Note, however, that in spite of these poor constraints, the variance contribution $\Delta GC$ of unresolved GCs to the SBF-signal is quite well determined: the uncertainties in $\sigma$ and TOMs for the Hydra giants translate to rather small uncertainties below 0.05 mag for $\Delta GC$, see Sect.~\ref{sbfmeas}, due to the bivariance between $\sigma$ and TOM.\\
Regarding the Centaurus cluster giant NGC 4709, we would like to add a note: our colour estimate of $(V-I)=$ 1.35 mag for NGC 4709 is 0.13$-$0.15 mag redder than that of Tonry et al.~(\cite{Tonry01}) or Jensen et al.~(\cite{Jensen01}) for the same galaxy. Adopting the Tonry colour results in an 0.6 mag higher distance of 33.1 mag. This gives a stronger disagreement with the GCLF-distance for that galaxy, but at the same time a nice agreement with the SBF-distance of NGC 4696. Carefully re-analysing our photometry shows that the sky level is the parameter that most influences the colour, given that the halo of NGC 4709 extends to almost the image limits. Systematic errors in the photometric calibration are unlikely as the mean SBF-distance to the five other galaxies in the field of NGC 4709 agree to within 2\% with the mean of the entire sample. The amount of large scale flat field variations can theoretically account for a colour shift of up to $\simeq$ 0.10 mag. However, we find the particular sky value adopted for NGC 4709 to be consistent with the large scale flat field variations derived in other fields not dominated by one major galaxy, indicating that our colour value should not be off by more than about 0.05 mag from the real colour. For a detailed discussion of this we refer to a paper in preparation about the photometric parameters of the Centaurus giants and their globular cluster systems (Hilker et al. in preparation).\vspace{0.1cm}\\
Summarising the discussion on possible systematic effects within our Hydra and Centaurus data,
we find an upper limit of about 0.1 mag for too low Hydra distances and too high Centaurus distances with bluer
colour. We find that Malmquist like biases are unlikely to cause a bias larger than 
0.05 mag.
\subsection{Consequence of a higher Hydra and lower Centaurus 
distance for the GA model}
\label{consGA}
Due to the difference between our distance estimates to Hydra and Centaurus to those of other authors we check the peculiar velocity predicted for a 15\% larger Hydra and 15\% smaller Centaurus distance within the flow model of Tonry et al.(\cite{Tonry00}), see Fig.~\ref{GAraddif}.\\  
The expected CMB radial velocity for a higher Hydra distance
remains well below the actually measured ones for GA distances between 43 and 47 Mpc. This is because a 15\% higher distance places Hydra behind the GA, resulting in negative peculiar velocities. However, for a GA distance of 49 Mpc and a projected position within about 10$\degr$ of the Hydra cluster, the expected CMB radial velocity of a 15\% more distant Hydra cluster matches the observations. This GA distance is 15\% higher than determined by T00, different at 2$\sigma$ significance. Note that even for this larger assumed Hydra distance, the Hydra peculiar velocity is still significant and requires a strong gravitational pull from behind, consistent with a GA in closer projected position to Hydra than to Centaurus. When in addition assuming a 15\% lower Centaurus-distance, the Hydra and Centaurus peculiar velocities become comparable. This can be explained by a 49 Mpc distant GA at approximately equal projected distance between Hydra and Centaurus, but also by a common motion towards a more distant attractor.\\
Although a check of the predicted peculiar velocities using literature distances is a valuable exercise, we note that it is difficult to imagine that any systematic effect biases our SBF-distances by 15\% (0.3 mag) in {\it opposite} directions for Hydra and Centaurus, see Sect.~\ref{syseff}. That is, the significant {\it difference} between the Hydra and Centaurus peculiar velocity should be less affected by systematics than their absolute distances. This is especially true given that both data sets were observed with the same instrument.
\section{Summary and conclusions}
We have presented $I$-band SBF-measurements for 16 early type Hydra cluster galaxies in
the magnitude range $10<V<18.5$ mag, including 13 dwarfs. 
The measurements are based on deep photometric data obtained
with VLT FORS1 in the $I$-band in 7 fields with a seeing between 0.6 and 0.7$\arcsec$. 
The following results
were obtained:\vspace{0.15cm}\\
\indent 1. The mean SBF-distance of the investigated Hydra galaxies is 41.2
$\pm$ 1.4 Mpc (33.07 $\pm$ 0.07 mag).
Together with the same kind of SBF data presented for the Centaurus cluster by Mieske \& Hilker
(\cite{Mieske03b}) and revised in this paper, this gives a relative distance between Centaurus and
Hydra of 
$(m-M)_{\rm Cen}-(m-M)_{\rm Hyd}=0.21 \pm 0.11$ mag, or 4.1 $\pm$ 2.4 Mpc.\\
\indent 2. In the CMB rest-frame and assuming $H_{\rm 0}=72\pm 4$ km~s$^{-1}$ Mpc$^{-1}$,
the distance obtained for the Hydra cluster yields a high positive peculiar velocity of 1225
$\pm$ 235 km~s$^{-1}$. Together with the  lower peculiar velocity of the Centaurus
cluster (210 $\pm$ 295 km~s$^{-1}$) and allowing for a thermal velocity error component of 200 km~s$^{-1}$, this rules out a common flow velocity for both clusters at 98\% confidence. We find that the $9\times 10^{15} M_{\sun}$``Great Attractor'' from the flow study of Tonry et al.~(\cite{Tonry00}) at a distance of 43 $\pm$ 3 Mpc can explain the observed peculiar velocities if shifted about 15$\degr$ towards the Hydra cluster position. 
Our results are inconsistent at 94\% confidence with a scenario where the Centaurus cluster is the GA. The difference between the mass of the Centaurus cluster as derived from X-ray observations and that proposed for the Great Attractor is about a factor of 30, making that hypothesis even more unlikely. The possibility of a large Hydra peculiar velocity due to infall along a filament is inconsistent with our data. The idea of a net angular momentum of the Hydra-Centaurus system in addition to the cosmic thermal velocity field is found to be inadequate due to the large time-scales and distances involved.\\
\indent 3. The Hydra cluster SBF-distance derived by us is about 15\% lower than the mean of distances published in the last five years, while the estimated
Centaurus distances agree well with FP-distances, but are about 15\% higher than previous SBF-estimates. 
Several possible reasons for
these differences are discussed, for example peculiarities in the stellar 
population of the bluest galaxies in our sample that might bias our 
sample towards low (Hydra) or high (Centaurus)
distances by up to 0.1 mag, or peculiarities in the stellar
population of the brightest 
cluster galaxies that may bias the distance measurements for IR-SBF or Fundamental Plane
measurements by up to 0.2 mag. We find that also for a 15\% higher Hydra distance, its peculiar velocity is substantial and consistent with a massive attractor in close projection to and slightly beyond Hydra.\\
\indent 4. We cannot place lower limits on the Hydra cluster depth from our data.
The scatter of the SBF-distances around their mean is slightly below
the mean single measurement uncertainty. With 95\% confidence, a radial
extension of more than $\pm$ 3.5 Mpc is ruled out by the data. This upper limit corresponds
to a cigar shape with a 3-4 times longer radial than tangential extension.\\\\
It is concluded that an updated modelling of the local peculiar velocity field
might become necessary in the light of more and more high quality data arriving
for distance estimates to nearby galaxies. In order to better restrict partially degenerate Great Attractor parameters like its mass and distance, a recalculation of the local flow model with
updated distance information over a much larger area than covered by us is necessary. The discovery of very massive galaxy clusters in the Zone of Avoidance in the last few years at substantial angular separation from the Hydra-Centaurus region supports the impression that there is not one single ``Great Attractor'', but rather several substructures within a generally overdense filamentary region of the nearby universe.
\noindent 
\label{conclusions}
\acknowledgements
We thank the anonymous referee for his constructive comments which helped to
improve the paper. We thank John P. Blakeslee and John L. Tonry for providing 
us with details of the results from their SBF-survey. In addition we thank
John P. Blakeslee for fruitful discussions. We are grateful to 
Russell J. Smith for information
and discussion on the results of the SMAC survey.
SM was supported by DAAD PhD grant Kennziffer D/01/35298 and DFG project HI
855/1. LI acknowledges support from FONDAP ``Center for Astrophysics''.
The authors would like to thank the ESO user support group and the ESO science operation for having 
carried out the programme in service mode.

\enddocument